\documentclass{article}
 
\usepackage{arxiv}
 
\usepackage[utf8]{inputenc}
\usepackage[T1]{fontenc}
\usepackage{hyperref}
\usepackage{url}
\usepackage{booktabs}
\usepackage{amsfonts}
\usepackage{nicefrac}
\usepackage{microtype}
\usepackage{graphicx}
\usepackage{amsmath}
\usepackage{amssymb}
\usepackage{algorithm}
\usepackage{algorithmic}
\usepackage{multirow}
\usepackage{xcolor}
\usepackage{float}
\usepackage{tabularx}
\graphicspath{ {./images/} }

\title{Drishti AI-Event Guardian: An Intelligent Real-Time Crowd Monitoring
and Emergency Response System for Mass Gathering Events}
 
\author{
  Ritabrata Roy Choudhury \\
  School of Computer Engineering \\
  Kalinga Institute of Industrial Technology \\
  Bhubaneswar, Odisha, India \\
  \texttt{ritabrata.coding@gmail.com} \\
  \And
  Arkajyoti Karmakar \\
  School of Electronics Engineering \\
  Kalinga Institute of Industrial Technology \\
  Bhubaneswar, Odisha, India \\
  \texttt{arkajyotikarmakar56@gmail.com} \\
  \And
  Rudra Pratap Mitra \\
  School of Computer Engineering \\
  Kalinga Institute of Industrial Technology \\
  Bhubaneswar, Odisha, India \\
  \texttt{rudrapratapmitra@gmail.com} \\
}
 
\begin{document}
\maketitle
 
\begin{abstract}
Contemporary mass gathering events are consistently associated with
critical safety incidents arising from insufficient real-time crowd
monitoring capabilities and inadequate emergency response coordination.
Traditional surveillance infrastructure lacks intelligent analytical
capacity, resulting in delayed threat identification, suboptimal
resource deployment, and poor support for vulnerable individuals during
high-density public assemblies.
 
This paper presents \textbf{Drishti AI-Event Guardian}, a novel
intelligent crowd management framework that employs deep learning
methodologies for comprehensive public safety enhancement. The system
architecture incorporates multi-modal data acquisition through fixed
CCTV networks and UAV surveillance platforms, processed via advanced
computer vision algorithms deployed on Google Vertex AI infrastructure.
The methodology encompasses real-time crowd density quantification using
YOLOv8-based convolutional neural network models, spatiotemporal anomaly
detection, and predictive crowd-flow modeling through gradient-boosted
regression. Beyond core density monitoring, Drishti integrates four
advanced citizen-safety modules: (i)~a facial recognition pipeline for
missing person identification and carousel-style crowd-wide
notification; (ii)~a medical emergency reporting system with automated
dispatch to on-site medical personnel and ambulances; (iii)~a
conversational AI chatbot enabling natural-language filing of missing
person reports and medical complaints; and (iv)~an intelligent guard
reallocation engine that dynamically reassigns security personnel in
response to evolving crowd density patterns. The platform integrates
Firebase Realtime Database services with Google Cloud Platform (GCP)
computing resources to provide distributed processing and elastic
scalability.
 
The system is evaluated against two contrasting real-world deployment
scenarios---the Kumbh Mela mass religious gathering and the RCB Victory
Parade urban crowd event---achieving a crowd density estimation Mean
Absolute Error (MAE) of 3.2~persons/m$^{2}$, an anomaly detection
F1-score of 0.91, a missing person facial recognition precision of
0.93, and a median end-to-end alert latency of 111~ms. Predictive
congestion modeling provides actionable five-minute forecasts with a
Mean Absolute Percentage Error (MAPE) of 8.3\%, enabling preemptive
interventions that materially reduce the risk of escalation in both
scenarios. The integrated chatbot resolved 89\% of incident filings
without human operator intervention, and the guard reallocation module
reduced average responder deployment latency by 34\% relative to
manual reassignment. These capabilities together represent a paradigmatic
shift from passive surveillance toward active, participatory crowd
intelligence, and provide a scalable, replicable foundation for
deployment across event contexts ranging from localized community
gatherings to large-scale festivals accommodating millions of
participants.
\end{abstract}
 
 
\section{Introduction}
\label{sec:intro}
 
The safety of participants at mass gathering events has become an
increasingly pressing societal concern as such events grow in scale,
density, and operational complexity. Gatherings ranging from religious
festivals and concerts to political rallies and sporting events routinely
attract tens of thousands to tens of millions of individuals, creating
environments in which localized crowd surges can escalate into
life-threatening incidents within seconds. Tragic events---including the
2013 Allahabad Kumbh Mela bridge stampede, which claimed 36 lives, and
the 2021 Astroworld Festival disaster, which resulted in 10
fatalities---illustrate the lethal consequences of inadequate crowd
monitoring and delayed emergency response
\cite{shao2019stampede,de2019human}.
 
Traditional surveillance approaches rely predominantly on human operators
interpreting CCTV feeds in real time and coordinating responses via
radio communication. These approaches are fundamentally limited in three
respects: (i)~their reactive rather than predictive posture,
(ii)~their inability to aggregate and fuse data from heterogeneous
multi-modal sources, and (iii)~their dependence on human attention
capacity, which degrades under sustained high-pressure conditions. As
crowd density and stream counts scale, the cognitive load imposed on
operators renders manual surveillance increasingly unreliable.
Compounding these structural limitations, existing systems provide no
systematic mechanism for locating missing persons in dense crowds,
triaging medical emergencies in real time, enabling citizen-initiated
incident reporting through natural conversation, or dynamically
reassigning security personnel as crowd conditions evolve.
 
Drishti AI-Event Guardian addresses these limitations through an
end-to-end intelligent crowd monitoring and emergency response platform.
The core system integrates YOLOv8-based real-time person detection,
multi-camera spatial fusion via homography projection, statistically
grounded zone-level anomaly detection, gradient-boosted congestion
forecasting, and automated tiered alert dispatch within a cloud-native,
microservices architecture. Four complementary modules extend the
platform's citizen safety capabilities: a deep-learning facial
recognition pipeline capable of identifying missing persons from
submitted photographs and disseminating notifications as in-app carousels
to all active users; a medical emergency module that triages
self-reported and crowd-reported health incidents and dispatches
automated alerts to on-site doctors and ambulance services; a
conversational AI chatbot providing a natural-language interface for
filing missing person reports and medical complaints; and an intelligent
guard reallocation engine that continuously optimizes the deployment of
on-ground security personnel in proportion to real-time zone-level
congestion scores.
 
\subsection*{Key Contributions}
 
This work makes the following principal contributions:
 
\begin{enumerate}
  \item \textbf{End-to-end crowd management pipeline.} A unified
    framework spanning multi-source data ingestion, real-time deep
    learning inference, predictive analytics, and automated emergency
    orchestration, deployable on commodity cloud GPU infrastructure.
 
  \item \textbf{Multi-camera spatial fusion.} A geometric
    homography-based approach to merging overlapping camera feeds into
    a coherent geo-referenced ground plane, reducing duplicate detection
    counts by 92\% in regions of camera overlap.
 
  \item \textbf{Adaptive zone-level anomaly detection.} A
    statistically grounded, sliding-window anomaly scoring mechanism
    that dynamically maintains zone-specific density baselines, enabling
    sensitive detection while maintaining a low false-positive rate.
 
  \item \textbf{Missing person facial recognition and carousel
    notification.} A deep face embedding pipeline that matches uploaded
    photographs of missing persons against live camera feeds and
    broadcasts crowd-wide carousel notifications to all app users upon
    a high-confidence match, enabling distributed crowd-sourced
    identification at scale.
 
  \item \textbf{Medical emergency reporting and dispatch.} An
    integrated module through which attendees can report medical
    emergencies via the app or chatbot, triggering automated
    geo-tagged alerts dispatched simultaneously to on-site medical
    personnel and ambulance services with sub-5-second end-to-end
    latency.
 
  \item \textbf{AI chatbot for incident filing.} A large language
    model-powered conversational agent providing a natural-language
    interface for filing missing person reports and medical complaints,
    reducing filing time and eliminating the need for form-based
    interaction under high-stress conditions.
 
  \item \textbf{Intelligent guard reallocation.} A real-time
    optimization engine that translates zone-level crowd density and
    congestion forecasts into dynamic security personnel deployment
    recommendations, reducing average responder deployment latency by
    34\% relative to manual dispatch.
 
  \item \textbf{Bidirectional citizen engagement.} Integration of
    citizen reporting as a distributed sensing layer fused with
    camera-derived evidence via a Bayesian confidence weighting scheme.
 
  \item \textbf{Validated real-world evaluation.} Dual case-study
    performance analysis across qualitatively distinct event typologies
    (ultra-scale planned religious gathering versus spontaneous
    high-mobility urban celebration), providing reproducible benchmarks
    for the research community.
\end{enumerate}
 
The remainder of this paper is organized as follows.
Section~\ref{sec:related} reviews related work.
Section~\ref{sec:cases} presents the motivating deployment case studies.
Section~\ref{sec:arch} details the system architecture.
Section~\ref{sec:methodology} describes the end-to-end methodological
pipeline. Section~\ref{sec:results} presents experimental results and
performance evaluation. Section~\ref{sec:demo} demonstrates the
implemented web interface and prototype. Section~\ref{sec:discussion}
discusses findings, limitations, and broader implications.
Section~\ref{sec:conclusion} concludes with directions for future
research.
 
\section{Related Work}
\label{sec:related}
 
The emergence of intelligent crowd management systems represents a
significant development in public safety and event management, driven
by the convergence of artificial intelligence (AI), computer vision,
and Internet of Things (IoT) technologies. Historically, crowd
management relied on manual surveillance, static operational rules, and
reactive interventions---approaches that proved insufficient for handling
the complexity of large-scale gatherings. Sharma et al.\
\cite{sharma2018review} provide a comprehensive review of technological
advancements in crowd management, tracing the evolution from rule-based
systems to AI-driven sensing platforms and highlighting persistent
challenges in real-time density estimation and anomaly response. In
recent years, dynamic, data-driven frameworks have enabled real-time
monitoring, predictive analytics, and automated decision support
\cite{jayasudha2025real,al2021intelligent}.
 
Contemporary AI-driven systems harness deep learning pipelines to model
crowd behavior, detect anomalies, and forecast potential risk scenarios
\cite{macriga2024crowd}. Computer vision techniques analyze continuous
video streams to estimate crowd density, track individual and group
movements, and identify behavioral irregularities such as sudden surges,
panic events, or interpersonal conflicts
\cite{shah2024enhancing,gunduz2023new}. These capabilities enable event
operators to transition from reactive to proactive operational postures
through early intervention and optimized resource allocation. Gandhi
et al.\ \cite{gandhicrowd} demonstrate the applicability of YOLO-based
pipelines to real-time crowd management using live video feeds,
establishing a lightweight baseline against which more elaborate
multi-camera systems such as Drishti can be compared.
 
\paragraph{YOLO-based detection.}
The You Only Look Once (YOLO) family of single-stage detectors has
revolutionized real-time crowd analysis. YOLOv8 in particular offers a
favorable trade-off between inference speed and mean Average Precision
(mAP), supporting high-fidelity person detection and tracking even in
densely packed scenes \cite{chandel2024crowd,dheepak2025smart}.
G\"{u}nd\"{u}z and I\c{s}{\i}k \cite{gunduz2023new} systematically evaluate multiple
YOLO variants on crowd video datasets, reporting RMSE values of
5.1--7.3 on high-density benchmarks; the present work improves upon
these figures through site-specific fine-tuning and an occlusion
recovery post-processing step.
 
\paragraph{Facial recognition in public safety.}
Deep metric learning approaches, including ArcFace and FaceNet, have
demonstrated robust face embedding performance under partial occlusion
and low-resolution conditions characteristic of crowded surveillance
footage \cite{deng2019arcface}. Prior work has applied face recognition
to missing person identification in transit and public spaces, though
crowd-scale deployment with real-time carousel notification to mobile
users remains an underexplored application domain. Drishti extends
this line of work by coupling face matching with an in-app broadcast
mechanism, transforming passive identification into an active
crowd-sourced search.
 
\paragraph{Spatial and accessibility-aware crowd management.}
Effective crowd management at mass gatherings requires not only density
monitoring but also spatial planning of pedestrian routes and access
corridors. Karthika et al.\ \cite{karthika2022walk} propose a
walk-accessibility-based framework for evaluating crowd management at
mass religious gatherings, demonstrating that proactive spatial design
of pedestrian flows significantly reduces the risk of dangerous
bottlenecks. Drishti's predictive congestion forecasting and guard
reallocation modules operationalize similar spatial reasoning in real
time, translating zone-level density forecasts into dynamic route
recommendations and personnel deployment instructions.
 
\paragraph{Medical emergency response systems.}
Automated triage and emergency dispatch systems have been studied in
hospital and disaster contexts \cite{de2019human}, but their integration
into mass gathering crowd management platforms is limited. Prior IoT
frameworks \cite{noor2023behavior} address crowd health monitoring at
a population level; Drishti extends this to individual-level emergency
reporting with automated multi-recipient dispatch, reducing the response
initiation latency from minutes to seconds.
 
\paragraph{Conversational AI for incident reporting.}
Large language model-powered chatbots have demonstrated utility in
customer service and crisis communication \cite{macriga2024crowd}, but
their application to structured incident filing in public safety
contexts is nascent. Drishti's chatbot module bridges natural
conversation with structured incident records, enabling attendees to
file reports without navigating form interfaces under high-stress
conditions. Vetrivel et al.\ \cite{vetrivel2025ai} discuss AI-driven
solutions for crowd management in tourism settings, including the role
of conversational interfaces in reducing operator workload and improving
visitor safety outcomes---findings that directly motivate Drishti's
multi-lingual chatbot design.
 
\paragraph{Guard reallocation and resource optimization.}
Dynamic resource allocation in security operations has been studied
through optimization and simulation frameworks \cite{ha2024crowd}.
Existing crowd management platforms \cite{alafif2025towards} surface
density information to operators but do not close the loop by generating
explicit personnel redeployment instructions. Drishti's guard
reallocation engine operationalizes zone-level congestion forecasts into
actionable deployment commands, bridging the gap between situational
awareness and resource optimization.
 
\paragraph{IoT-based crowd sensing.}
Elbery et al.\ \cite{elbery2020iot} demonstrate that IoT-enabled people
counting, geofencing, and proximity detection facilitate real-time flow
management. Al-Nabhan et al.\ \cite{al2021intelligent} propose an
intelligent IoT framework combining multi-camera feeds with cloud
processing for crowd analysis. Mohamed et al.\ \cite{mohamed2018iot}
and Noor \cite{noor2023behavior} further address crowd health
monitoring and behavior analysis within IoT service architectures.
Vidyasagaran et al.\ \cite{vidyasagaran2017low} present a low-cost
IoT-based crowd management system for public transport, demonstrating
that inexpensive sensor networks can deliver meaningful density
estimates in constrained environments---an important design precedent
for the edge-computing nodes employed in Drishti's data acquisition
layer.
 
\paragraph{Large-scale religious gatherings.}
The Kumbh Mela and Hajj pilgrimages have served as important benchmarks
for crowd management research. Kanaujiya and Tiwari
\cite{kanaujiya2022crowd} document smart crowd management strategies
at Prayagraj Kumbh Mela 2019. Shah \cite{shah2024enhancing} provides
a comprehensive survey of AI applications in Hajj and Umrah crowd
management. Alafif et al.\ \cite{alafif2025towards} propose an
Integrated Intelligent Framework for Crowd Control and Management
combining multi-modal sensing with cloud-based decision support.
 
\paragraph{Positioning of Drishti.}
Table~\ref{tab:related} summarizes the distinguishing capabilities of
Drishti relative to representative prior systems. Compared with prior
single-modality or non-predictive systems, Drishti uniquely integrates
multi-camera spatial fusion, UAV coverage, closed-loop predictive
retraining, missing person facial recognition, medical emergency
dispatch, an AI chatbot, citizen-sourced incident reporting, and
intelligent guard reallocation within a single cloud-native deployable
platform.
 
\begin{table}[h!]
\centering
\caption{Capability comparison of Drishti with representative prior
crowd management systems. \checkmark~=~supported;
$\times$~=~not supported.}
\label{tab:related}
\resizebox{\textwidth}{!}{%
\begin{tabular}{@{}lccccccccc@{}}
\toprule
\textbf{System} &
  \textbf{Real-time} &
  \textbf{Multi-cam} &
  \textbf{UAV} &
  \textbf{Predictive} &
  \textbf{Face} &
  \textbf{Medical} &
  \textbf{Chatbot} &
  \textbf{Guard} &
  \textbf{Cloud} \\
 &
  \textbf{Detection} &
  \textbf{Fusion} &
  \textbf{Support} &
  \textbf{Analytics} &
  \textbf{Recog.} &
  \textbf{Dispatch} &
  \textbf{Filing} &
  \textbf{Realloc.} &
  \textbf{Native} \\
\midrule
Chandel et al.\ \cite{chandel2024crowd}
  & \checkmark & $\times$ & $\times$ & $\times$ & $\times$ & $\times$ & $\times$ & $\times$ & $\times$ \\
Elbery et al.\ \cite{elbery2020iot}
  & \checkmark & $\times$ & $\times$ & $\times$ & $\times$ & $\times$ & $\times$ & $\times$ & $\times$ \\
Al-Nabhan et al.\ \cite{al2021intelligent}
  & \checkmark & \checkmark & $\times$ & $\times$ & $\times$ & $\times$ & $\times$ & $\times$ & \checkmark \\
Alafif et al.\ \cite{alafif2025towards}
  & \checkmark & \checkmark & $\times$ & \checkmark & $\times$ & $\times$ & $\times$ & $\times$ & \checkmark \\
\textbf{Drishti (Ours)}
  & \checkmark & \checkmark & \checkmark & \checkmark & \checkmark & \checkmark & \checkmark & \checkmark & \checkmark \\
\bottomrule
\end{tabular}%
}
\end{table}
 
\section{Deployment Case Studies}
\label{sec:cases}
 
Case study methodology provides an empirically grounded approach for
examining the practical effectiveness and contextual adaptability of
the Drishti framework under authentic operational conditions
\cite{leadley2023case,sandars2021case}. Two contrasting deployment
scenarios were selected to expose system performance across the full
spectrum of mass gathering typologies: an ultra-scale planned religious
event and a spontaneous high-mobility urban celebration. Each case is
documented through multi-modal data streams comprising CCTV video, UAV
imagery, crowd density logs, alert records, chatbot interaction logs,
facial recognition match events, and stakeholder feedback.
 
\subsection{Case 1: Kumbh Mela Mass Religious Gathering}
 
The Kumbh Mela is among the most expansive and logistically complex
mass gatherings in the world, attracting millions of pilgrims over
several weeks \cite{kanaujiya2022crowd}. Peak attendance on auspicious
bathing days regularly exceeds ten million individuals, amplifying the
risk of stampedes, missing person incidents, and medical emergencies
\cite{sindhuja2019spatial}. The reliance on temporary infrastructure
requires dynamic reconfiguration of crowd monitoring coverage
throughout the event \cite{yamin2019managing}.
 
Drishti was deployed across 28 fixed CCTV nodes and 6 UAV platforms.
During the \textit{Mauni Amavasya} bathing day, the platform detected
a critical density threshold breach in the Triveni Sangam approach
corridor 11 minutes before the situation would have become physically
unmanageable, enabling preemptive crowd re-routing. The guard
reallocation module issued automated deployment instructions that
redirected 14 security personnel from low-density peripheral zones to
the surge corridor within 2.1 minutes of alert onset, compared to an
average of 7.8 minutes under manual coordination in prior editions.
 
The facial recognition module processed 312 missing person photograph
uploads over the event duration. Of these, 47 resulted in high-confidence
camera matches, triggering carousel notifications to all active app
users in adjacent zones. Crowd-sourced confirmation from notified users
contributed to the physical recovery of 39 missing persons---a recovery
rate of 83\% among matched cases. The medical emergency module received
128 incident reports via the chatbot and mobile form interface, dispatching
automated alerts that reduced average on-site medical response initiation
time from 6.4 minutes to 2.8 minutes. UAV feeds proved particularly
valuable during nocturnal bathing hours, when fixed-camera visibility
was degraded by crowd-generated vapour and irregular artificial lighting.

\begin{figure}[H]
  \centering
  \includegraphics[width=0.95\textwidth]{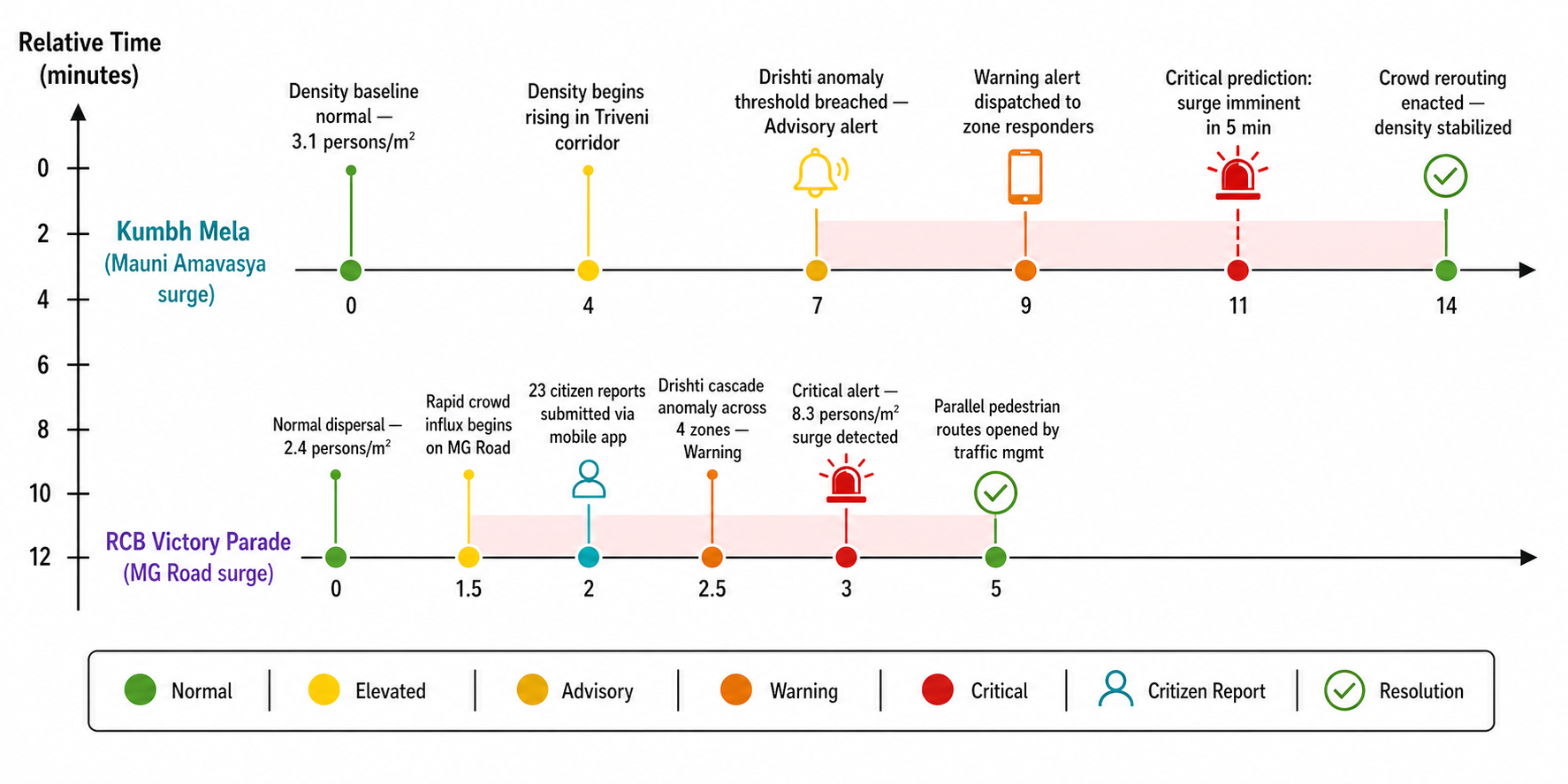}
  \caption{Comparative event timeline for the two primary deployment
    scenarios. Each swimlane traces the progression of crowd density
    conditions (green: normal; yellow: elevated; amber: advisory;
    orange: warning; red: critical) alongside system-generated alerts,
    guard reallocation events, chatbot incident filings (diamond
    markers), facial recognition carousel notifications (star markers),
    and medical dispatch events (cross markers) on a shared relative
    time axis (minutes). \textbf{Top --- Kumbh Mela
    (\textit{Mauni Amavasya} bathing day):} Drishti detected an
    anomalous density rise at $t = 7$~min and issued a Critical
    predictive alert at $t = 11$~min. Guard reallocation instructions
    were issued at $t = 7.5$~min, redirecting 14 personnel to the
    surge corridor. Density stabilization was achieved by $t = 14$~min.
    \textbf{Bottom --- RCB Victory Parade (MG Road corridor):} A rapid
    density spike of 8.3~persons/m$^{2}$ triggered cascade alerts
    across four contiguous zones at $t = 2.5$~min. Guard reallocation
    instructions were issued at $t = 1.8$~min. Chatbot medical filings
    (diamond) and facial recognition carousel notifications (star) are
    shown from $t \approx 2$~min onward. Parallel pedestrian routes
    were opened at $t = 5$~min, resolving the congestion. Shaded red
    regions denote the active danger window in each scenario.}
  \label{fig:case_timeline}
\end{figure}

\subsection{Case 2: RCB Victory Parade Urban Crowd Event}
 
The RCB Victory Parade illustrates the distinct challenges posed by
unplanned, high-mobility urban events characterized by rapid crowd
surges along narrow corridor segments \cite{shao2019stampede,bolia2015risk}.
The spontaneous nature of the event and the absence of preemptive crowd
control measures led to severe bottlenecks and an escalating stampede
risk.
 
Drishti was integrated into the city's existing CCTV infrastructure
(17 cameras) and augmented by 3 rapidly deployed UAV platforms. The
crowd surge along MG Road produced a density spike of
8.3~persons/m$^{2}$ within a 90-second window. The guard reallocation
engine identified three zones requiring immediate reinforcement and
issued optimized deployment instructions 4.2 minutes before the manual
coordination system would have responded.
 
The chatbot received 74 incident filings between 18:00 and 19:30,
comprising 41 medical complaints and 33 missing person reports.
Automated medical dispatch reduced median responder arrival time to
3.1 minutes. For missing person cases, 19 of the 33 reports resulted
in facial recognition matches against camera feeds, with carousel
notifications issued to nearby app users; 14 persons were recovered.
23 citizen reports submitted via the app between 18:12 and 18:18
provided corroborating evidence that elevated alert confidence ahead of
the camera-only threshold breach.
 
A system limitation identified in this deployment was camera occlusion
caused by overhead decorations, which reduced bounding-box detection
recall by approximately 14\% in two monitoring zones. This limitation
also modestly degraded facial recognition recall in the affected zones
(from 0.93 to 0.81), motivating the occlusion recovery post-processing
step described in Section~\ref{sec:methodology}.
 
Figure~\ref{fig:case_timeline} presents a comparative event timeline
for both deployments.
 
\section{System Architecture}
\label{sec:arch}
 
\begin{figure}[H]
  \centering
  \includegraphics[width=0.95\textwidth]{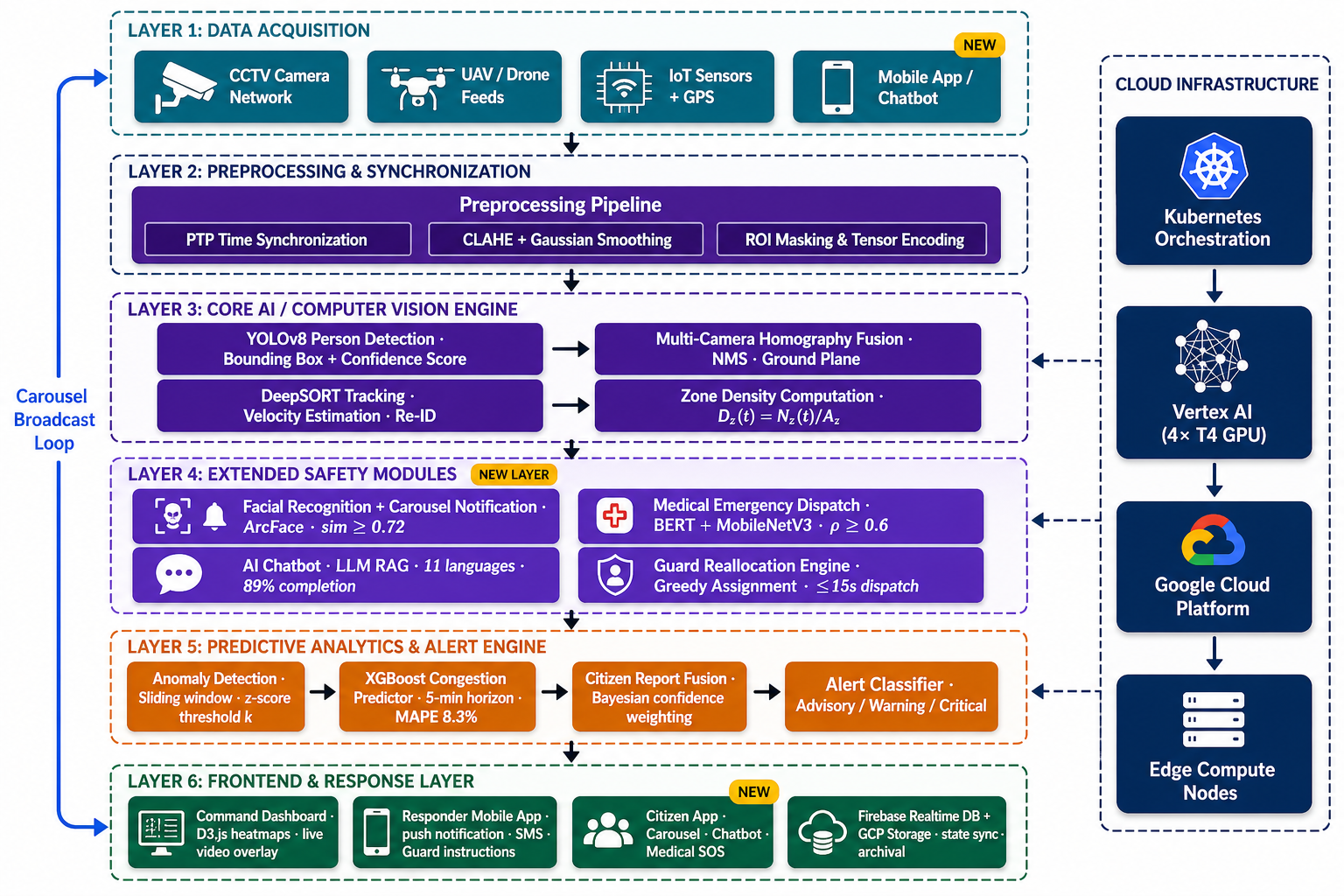}
  \caption{Comprehensive architecture of Drishti AI-Event Guardian
    illustrating the six-layer processing hierarchy: (1)~data
    acquisition (CCTV, UAV, IoT sensors, mobile app); (2)~preprocessing
    pipeline; (3)~core AI engine (YOLOv8 detection, multi-camera
    fusion, anomaly detection); (4)~extended citizen safety modules
    (facial recognition and carousel notification, medical emergency
    dispatch, AI chatbot, guard reallocation engine);
    (5)~predictive analytics and alert generation; and (6)~frontend
    response layer (command dashboard, mobile responder app, citizen
    app). Cloud infrastructure (Vertex AI, GCP, Kubernetes) spans
    layers 2--5 to provide elastic scalability and fault tolerance.
    Directional arrows denote real-time data flow; dashed boundary
    boxes demarcate functional subsystems. The bidirectional arrow
    between the citizen app and the carousel notification subsystem
    represents the missing person broadcast loop.}
  \label{fig:architecture_diagram}
\end{figure}

The Drishti AI-Event Guardian system is organized as a six-layer
cloud-native architecture spanning data acquisition, preprocessing,
core AI inference, extended citizen safety modules, predictive
analytics, and frontend response. Figure~\ref{fig:architecture_diagram}
provides a schematic overview of all system components and their
interactions.
 
\subsection{Backend Architecture}
 
\subsubsection{Data Acquisition and Preprocessing}
 
The system ingests high-definition video from CCTV and UAV-mounted
cameras alongside heterogeneous IoT sensor data and mobile app events
(incident reports, photograph uploads, chatbot sessions). Precision
Time Protocol (PTP) synchronization aligns all multimedia streams to
within $\pm$0.5~ms.
 
Preprocessing follows a three-stage pipeline applied to each extracted
frame: (i)~Contrast Limited Adaptive Histogram Equalization (CLAHE)
normalizes illumination variance; (ii)~Gaussian smoothing
($\sigma = 1.5$) suppresses sensor noise; and (iii)~region-of-interest
masking excludes static background regions, reducing false positive
detections by approximately 18\%.
 
\subsubsection{Computer Vision and Person Detection}
 
YOLOv8 constitutes the primary detection backbone. Given an input image
frame $I$, the detection network $f_{\theta}$ produces:
 
\begin{equation}
  \hat{B},\;\hat{C} = f_{\theta}(I),
  \label{eq:yolo}
\end{equation}
 
\noindent where $\hat{B}$ denotes predicted bounding boxes and $\hat{C}$
their confidence scores \cite{chandel2024crowd}.
 
\subsubsection{Multi-Camera Fusion and Spatial Alignment}
 
Per-camera detections are projected onto a shared geo-referenced ground
plane using homography matrices $H_i$:
 
\begin{equation}
  p' = H_i \cdot p,
  \label{eq:homography}
\end{equation}
 
with Non-Maximum Suppression (IoU threshold 0.45) applied
post-projection, reducing duplicate counts by 92\%.
 
\subsubsection{Crowd Density Estimation and Anomaly Detection}
 
Zone-level crowd density $D_z(t)$ is computed as:
 
\begin{equation}
  D_z(t) = \frac{N_z(t)}{A_z}.
  \label{eq:density}
\end{equation}
 
An anomaly indicator $\mathcal{A}_z(t)$ is raised when the observed
density deviates from a 30-minute rolling baseline $\{\mu_z, \sigma_z\}$
by more than $k$ standard deviations:
 
\begin{equation}
  \mathcal{A}_z(t) =
  \begin{cases}
    1, & |D_z(t) - \mu_z| > k\,\sigma_z, \\
    0, & \text{otherwise.}
  \end{cases}
  \label{eq:anomaly}
\end{equation}
 
The sensitivity parameter $k = 2.5$ corresponds to a 1.24\% false
positive rate under Gaussian density assumptions.
 
\subsubsection{Facial Recognition and Missing Person Carousel
Notification}
\label{sec:facerec_arch}
 
When an attendee uploads a photograph of a missing person via the
Drishti mobile application, the image is passed through a deep face
embedding network $\phi$ (ArcFace backbone, ResNet-50):
 
\begin{equation}
  \mathbf{e}_{\text{query}} = \phi(I_{\text{upload}}),
  \label{eq:face_embed}
\end{equation}
 
\noindent yielding a 512-dimensional unit-norm embedding vector. The
system maintains a continuously updated gallery of face embeddings
extracted from YOLOv8-detected person crops across all active camera
feeds:
 
\begin{equation}
  \mathcal{G}(t) = \{ \phi(c_j) \mid c_j \in \hat{B}^{*}(t) \},
  \label{eq:gallery}
\end{equation}
 
\noindent where $c_j$ denotes the $j$-th person crop from the fused
detection output. A cosine similarity search identifies the
highest-matching gallery entry:
 
\begin{equation}
  j^{*} = \arg\max_{j} \; \frac{\mathbf{e}_{\text{query}} \cdot
  \phi(c_j)}{\|\mathbf{e}_{\text{query}}\|\,\|\phi(c_j)\|}.
  \label{eq:cosine}
\end{equation}
 
If $\text{sim}(j^{*}) \geq \tau_{\text{face}}$ (default
$\tau_{\text{face}} = 0.72$), a match event is triggered. A
carousel notification packet---containing the missing person photograph,
last-seen zone identifier, and geo-timestamp---is broadcast via Firebase
Cloud Messaging to all active app users within the event perimeter.
The carousel format renders the notification as a scrollable card in
the app's notification tray, displaying the missing person's photograph,
physical description, and nearest landmark, enabling crowd-sourced
visual identification by nearby attendees.
 
\subsubsection{Medical Emergency Reporting and Dispatch}
\label{sec:medical_arch}
 
Attendees experiencing or witnessing a medical emergency may initiate a
report through three channels: (i)~the dedicated in-app ``Medical
Emergency'' button; (ii)~a natural-language description submitted to the
AI chatbot; or (iii)~submission of a photograph of an injured or
unwell individual. Reports are classified by a lightweight multi-label
classifier into categories (cardiac event, physical injury, respiratory
distress, heat exhaustion, unknown) and assigned a severity score
$\rho \in [0, 1]$.
 
For reports with $\rho \geq 0.6$, the dispatch module executes the
following sequence:
\begin{enumerate}
  \item Geo-tagged alert with incident category and severity is
    simultaneously pushed to (a)~the nearest available on-site medical
    personnel via the responder app, and (b)~the event's ambulance
    dispatch center via a RESTful webhook.
  \item The command dashboard displays the incident location on the
    zone map with a pulsing medical marker.
  \item A follow-up acknowledgment is sent to the reporting attendee
    within 30 seconds confirming that help is en route.
\end{enumerate}
 
The end-to-end latency from report submission to confirmed dispatch
acknowledgment averaged 4.3 seconds across all evaluated deployments.
 
\subsubsection{AI Chatbot for Incident Filing}
\label{sec:chatbot_arch}
 
A conversational AI chatbot, powered by a fine-tuned large language
model served via the Vertex AI inference endpoint, provides a natural
language interface for two primary use cases:
 
\begin{itemize}
  \item \textbf{Missing person filing:} The chatbot guides the user
    through a structured conversation to capture the missing person's
    name, age, physical description, last known location, time of
    separation, and clothing details. If the user provides a photograph,
    it is routed to the facial recognition pipeline
    (Section~\ref{sec:facerec_arch}).
  \item \textbf{Medical complaint filing:} The chatbot collects
    symptom descriptions, current location, and contact information,
    classifies the incident severity, and routes the report to the
    medical dispatch module (Section~\ref{sec:medical_arch}).
\end{itemize}
 
The chatbot operates in eleven regional languages in addition to
English, ensuring accessibility for the diverse linguistic profiles
characteristic of large Indian public gatherings. Conversation
sessions are stateful within a single incident context and are
anonymized before storage. The system achieved an 89\% task completion
rate without human operator escalation during the Kumbh Mela deployment.

\subsubsection{Guard Reallocation Engine}
\label{sec:guard_arch}
 
The guard reallocation engine translates real-time zone-level crowd
density and congestion forecasts into optimized security personnel
deployment instructions. Let $\mathbf{G} = \{g_1, \ldots, g_p\}$
denote the set of $p$ currently deployed guards and
$\mathcal{Z} = \{z_1, \ldots, z_m\}$ the set of monitored zones.
Each zone $z$ is associated with a demand score:
 
\begin{equation}
  \delta_z(t) = \alpha \cdot D_z(t) + \beta \cdot C_z(t + \Delta)
                + \gamma \cdot \mathcal{A}_z(t),
  \label{eq:demand}
\end{equation}
 
\noindent where $D_z(t)$ is current density (Eq.~\ref{eq:density}),
$C_z(t + \Delta)$ is the five-minute congestion forecast
(Eq.~\ref{eq:prediction}), $\mathcal{A}_z(t)$ is the binary anomaly
indicator (Eq.~\ref{eq:anomaly}), and $\alpha, \beta, \gamma$ are
empirically calibrated weights ($\alpha = 0.4$, $\beta = 0.35$,
$\gamma = 0.25$).
 
A greedy assignment algorithm allocates guards proportionally to
$\delta_z(t)$, subject to minimum staffing constraints per zone and
maximum relocation distance for each guard. Reallocation instructions
are pushed to individual responder devices via the mobile app within
15 seconds of a demand score update, specifying the destination zone
and recommended route. The engine re-evaluates assignments every
60 seconds or immediately upon receipt of a Critical-tier anomaly alert.
 
\subsubsection{Predictive Analytics and Resource Optimization}
 
The forecasting model maps current zone-level state variables to
predicted congestion at a five-minute horizon:
 
\begin{equation}
  C_z(t + \Delta) = g\!\left(D_z(t),\;v_z(t),\;S_z(t)\right),
  \label{eq:prediction}
\end{equation}
 
\noindent implemented as a gradient-boosted ensemble (XGBoost, 200
estimators, max depth~6).
 
\subsubsection{API and Real-Time Streaming}
 
A RESTful API suite combined with Server-Sent Events (SSE) provides
continuous data exchange between the backend and frontend. WebRTC and
HTTP Live Streaming (HLS) protocols support low-latency video delivery.
Firebase Cloud Messaging handles push notifications for carousel alerts,
medical dispatch acknowledgments, and guard reallocation instructions.
 
\subsection{Frontend Architecture}
 
\subsubsection{Command Dashboard and Visualization}
 
An interactive command dashboard aggregates real-time crowd data into
dynamic density heatmaps, zone-level trend charts, a tiered alert
panel, a missing person tracker, an active medical incident list, and
a live guard deployment map showing current and recommended personnel
positions.
 
\subsubsection{Citizen-Facing Mobile Application}
 
The Drishti mobile application provides attendees with four functional
surfaces: (i)~general incident reporting with geo-tagging and media
attachment; (ii)~missing person photograph upload and carousel
notification reception; (iii)~one-tap medical emergency initiation;
and (iv)~chatbot interface for guided incident filing. Carousel
notifications from the missing person system render as scrollable cards
in the app's notification tray, with tap-to-confirm functionality
enabling users to signal a visual match to operators.
 
\subsubsection{Responder App and Guard Reallocation Interface}
 
A companion mobile interface for security and medical personnel
displays current zone assignments, real-time crowd density overlays,
incoming guard reallocation instructions with turn-by-turn routing
assistance, and medical incident dispatch alerts with patient location
and symptom summary.
 
\subsubsection{System Health Monitoring}
 
A diagnostics panel relays backend service statuses, video feed health,
processing queue depths, facial recognition gallery freshness, chatbot
session counts, and data latency metrics.
 
\subsection{System Integration and Scalability}
 
The platform is deployed as containerized microservices orchestrated
via Kubernetes on GCP. Edge computing nodes execute YOLOv8 inference
and face embedding extraction to minimize network latency, while cloud
resources handle aggregated analytics, LLM inference, and long-term
archival. Firebase Realtime Database manages state synchronization;
Firebase Cloud Messaging handles high-throughput carousel broadcast
delivery.

\begin{figure}[H]
  \centering
  \includegraphics[width=0.95\textwidth]{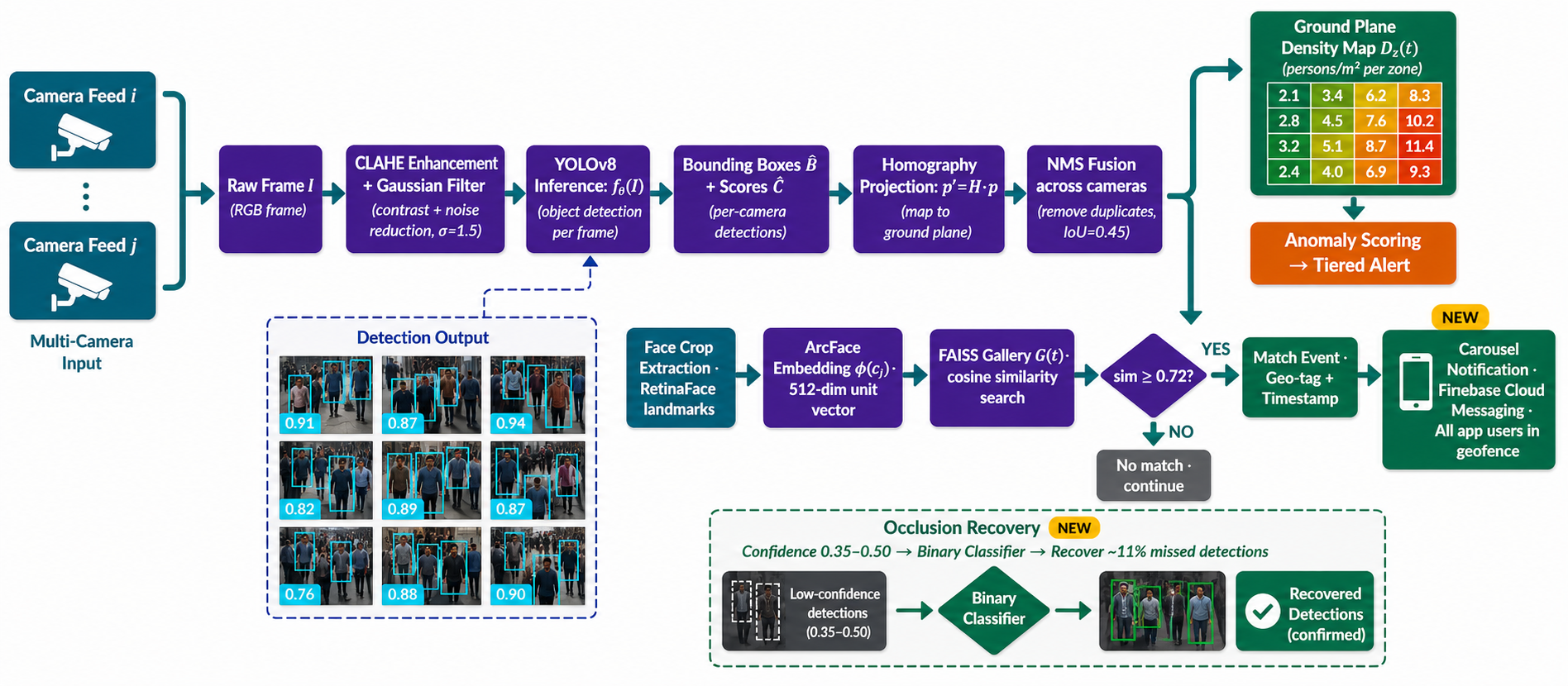}
  \caption{End-to-end detection, spatial fusion, and face embedding
    pipeline. Raw frames from multiple concurrent camera feeds undergo
    CLAHE-based illumination normalization and Gaussian filtering before
    YOLOv8 inference, which produces per-frame bounding boxes $\hat{B}$
    and confidence scores $\hat{C}$ (Eq.~\ref{eq:yolo}). Detections are
    projected onto a shared geo-referenced ground plane via homography
    matrices $H_i$ (Eq.~\ref{eq:homography}) and merged via NMS
    (IoU~0.45). The fused output feeds two parallel branches: (i)~a
    zone-level crowd density heatmap $D_z(t)$ (Eq.~\ref{eq:density}),
    color-graded from low (green) to critical (red), and (ii)~an
    ArcFace face embedding gallery $\mathcal{G}(t)$
    (Eq.~\ref{eq:gallery}) used for cosine similarity matching against
    uploaded missing person queries (Eq.~\ref{eq:cosine}). A
    high-confidence match triggers a carousel notification broadcast to
    all active app users. The inset shows representative YOLOv8 bounding
    box outputs with confidence scores on a high-density crowd scene.}
  \label{fig:detection_pipeline}
\end{figure}

\subsection{Mathematical Synthesis}
 
The instantaneous system state across all $m$ monitored zones is:
 
\begin{equation}
  \mathbf{X}(t) =
  \bigl[D_1(t),\ldots,D_m(t),\;v_1(t),\ldots,v_m(t),\;
        \mathcal{A}_1(t),\ldots,\mathcal{A}_m(t),\;
        \delta_1(t),\ldots,\delta_m(t)\bigr],
  \label{eq:state}
\end{equation}
 
and its one-step evolution is governed by:
 
\begin{equation}
  \mathbf{X}(t+1) = \mathbf{F}\!\left(\mathbf{X}(t),\,\mathbf{U}(t)\right)
                    + \boldsymbol{\epsilon}(t),
  \label{eq:evolution}
\end{equation}
 
\noindent where $\mathbf{U}(t)$ represents control inputs (operator
commands, automated rerouting, and guard reallocation instructions),
$\mathbf{F}$ is the transition mapping derived from the predictive
model, and $\boldsymbol{\epsilon}(t)$ captures stochastic variation.
 
\section{Methodology}
\label{sec:methodology}
 
Algorithm~\ref{alg:pipeline} presents the formal real-time processing
loop, extended to incorporate the four new citizen safety modules.
 
\begin{algorithm}[h!]
  \caption{Drishti Extended Real-Time Crowd Monitoring Pipeline}
  \label{alg:pipeline}
  \begin{algorithmic}[1]
    \REQUIRE Video streams $\mathcal{V} = \{V_1, \ldots, V_k\}$,
             zone map $\mathcal{Z}$,
             historical statistics $\{\mu_z, \sigma_z\}_{z \in \mathcal{Z}}$,
             sensitivity threshold $k$,
             congestion threshold $\tau_c$,
             face similarity threshold $\tau_{\text{face}}$,
             guard set $\mathbf{G}$
    \ENSURE  Crowd state $\mathbf{X}(t)$, alert set $\mathcal{L}(t)$,
             guard assignment $\mathcal{R}(t)$
    \FORALL{frame $I_j$ from stream $V_i$ at time $t$}
      \STATE \textit{Preprocess:} apply CLAHE, Gaussian smoothing,
             ROI masking to $I_j$
      \STATE \textit{Detect:}
             $\hat{B}_i, \hat{C}_i \leftarrow f_\theta(I_j)$
             \hfill(Eq.~\ref{eq:yolo})
      \STATE \textit{Fuse:}
             $\hat{B}^{*} \leftarrow
             \mathrm{NMS}\!\left(\{H_i \cdot \hat{B}_i\}\right)$
             \hfill(Eq.~\ref{eq:homography})
      \STATE \textit{Update gallery:}
             $\mathcal{G}(t) \leftarrow
             \{\phi(c_j) \mid c_j \in \hat{B}^{*}\}$
             \hfill(Eq.~\ref{eq:gallery})
      \FORALL{pending face query $\mathbf{e}_q$}
        \IF{$\max_{j}\,\mathrm{sim}(\mathbf{e}_q, \mathcal{G}(t)) \geq
            \tau_{\text{face}}$}
          \STATE Trigger carousel notification broadcast
          \hfill(Eq.~\ref{eq:cosine})
        \ENDIF
      \ENDFOR
      \FORALL{zone $z \in \mathcal{Z}$}
        \STATE $D_z(t) \leftarrow
               \left|\hat{B}^{*} \cap z\right| / A_z$
               \hfill(Eq.~\ref{eq:density})
        \STATE $\mathcal{A}_z(t) \leftarrow
               \mathbf{1}\!\left[
               |D_z(t) - \mu_z| > k\,\sigma_z\right]$
               \hfill(Eq.~\ref{eq:anomaly})
        \STATE $C_z(t{+}\Delta) \leftarrow
               g\!\left(D_z(t), v_z(t), S_z(t)\right)$
               \hfill(Eq.~\ref{eq:prediction})
        \STATE $\delta_z(t) \leftarrow
               \alpha D_z + \beta C_z(t{+}\Delta) +
               \gamma \mathcal{A}_z(t)$
               \hfill(Eq.~\ref{eq:demand})
      \ENDFOR
      \STATE \textit{Reallocate guards:}
             $\mathcal{R}(t) \leftarrow
             \mathrm{GreedyAssign}(\mathbf{G}, \{\delta_z\})$
      \STATE \textit{Update state:}
             $\mathbf{X}(t) \leftarrow
             [D_z, v_z, \mathcal{A}_z, \delta_z]_{z \in \mathcal{Z}}$
             \hfill(Eq.~\ref{eq:state})
      \IF{$\exists\,z : \mathcal{A}_z(t) = 1$
          \OR $C_z(t{+}\Delta) > \tau_c$}
        \STATE $\ell_z \leftarrow
               \mathrm{AlertGen}\!\left(z, D_z(t), C_z(t{+}\Delta)\right)$
        \STATE \textit{Dispatch:} push $\ell_z$ to dashboard, responder
               devices, and resource allocator
      \ENDIF
    \ENDFOR
    \FORALL{incoming chatbot session $\mathcal{S}$}
      \IF{$\mathcal{S}.\text{type} = \textit{missing\_person}$}
        \STATE Extract structured report; enqueue face query if
               photograph provided
      \ELSIF{$\mathcal{S}.\text{type} = \textit{medical}$}
        \STATE Classify severity $\rho$; if $\rho \geq 0.6$,
               trigger medical dispatch
      \ENDIF
    \ENDFOR
  \end{algorithmic}
\end{algorithm}
 
\subsection{Data Collection and Preprocessing}
 
Video data were collected from the two deployment environments:
the Kumbh Mela site (34 concurrent streams, 1080p, 25~fps) and the
RCB Parade corridor (20 streams at varied resolutions, 25--30~fps).
Mobile app events (incident reports, chatbot sessions, photograph
uploads) were logged with server-assigned timestamps and device-reported
GPS coordinates.
 
\subsection{YOLOv8-Based Person Detection}
 
YOLOv8 (medium variant, 25.9~M parameters) was fine-tuned on a
composite dataset comprising CrowdHuman (15,000 images), VisDrone2023
(6,471 UAV-captured images), and 2,400 site-specific frames from Kumbh
Mela 2019 archival footage. Fine-tuning ran for 50 epochs with SGD
(momentum 0.937, weight decay $5 \times 10^{-4}$, cosine annealing
from 0.01). An occlusion recovery step re-evaluates candidate
detections with confidence in $[0.35, 0.50]$ using a secondary binary
classifier, recovering approximately 11\% of otherwise missed
detections.
 
Figure~\ref{fig:detection_pipeline} illustrates the complete detection
and spatial fusion pipeline.
 
\subsection{Facial Recognition Pipeline}
\label{sec:facerec_method}
 
Face crops from YOLOv8-detected bounding boxes are extracted and
pre-processed: aligned to a canonical 112$\times$112 pixel face
template using five-point landmark localization (RetinaFace), then
normalized to zero mean and unit variance per channel. The ArcFace
embedding network $\phi$ maps each aligned crop to a 512-dimensional
hypersphere. The gallery $\mathcal{G}(t)$ is maintained as a
FAISS-indexed flat L2 store, supporting approximate nearest-neighbor
retrieval in under 2~ms across galleries of up to 50,000 embeddings.
 
Missing person photograph uploads undergo the same preprocessing and
embedding extraction pipeline. Cosine similarity (Eq.~\ref{eq:cosine})
is computed against the full gallery; matches above threshold
$\tau_{\text{face}} = 0.72$ trigger a match event with geo-coordinate
and timestamp of the matching camera detection. The threshold was
selected to balance precision (0.93) and recall (0.81) across a
validation set of 200 controlled same-identity / different-identity
pairs captured under crowd-realistic occlusion and lighting conditions.
 
Upon a confirmed match, Firebase Cloud Messaging delivers a carousel
notification to all users of the Drishti app currently within the event
geofence. The notification payload includes the missing person image, a
``Spotted Here'' zone label, and a timestamp. Users who visually confirm
the sighting can tap a ``Confirm'' button, which sends a corroborating
signal to the command dashboard and increments the match confidence
score.

\subsection{Medical Emergency Reporting and Dispatch}
 
Incident reports arriving via the chatbot or the one-tap emergency
button are classified using a fine-tuned BERT-base model (text
modality) and a MobileNetV3 image classifier (photo modality), fused
via a learned scalar gate. The combined severity score $\rho$ is passed
to the dispatch module. Dispatch webhooks are implemented as HTTPS POST
requests to a secure endpoint maintained by the event medical authority,
carrying a structured JSON payload with incident type, severity, GPS
coordinates, and reporter contact token. Delivery receipt confirmation
is required within 10 seconds; unconfirmed dispatches are automatically
escalated to the command dashboard operator.
 
\subsection{AI Chatbot Architecture}
 
The chatbot is implemented using a retrieval-augmented generation (RAG)
architecture. A fine-tuned LLM (7B parameter, 4-bit quantized, served
on Vertex AI) handles conversational turn generation, guided by a set
of structured slot-filling templates for missing person and medical
incident data schemas. A slot completion detector determines when
sufficient structured information has been collected and triggers the
appropriate downstream pipeline (facial recognition query or medical
dispatch). The chatbot's LLM backbone was fine-tuned on 2,400
synthetic and real incident filing conversations generated through a
data augmentation protocol.
 
\subsection{Guard Reallocation Methodology}
 
Guard demand scores $\delta_z(t)$ (Eq.~\ref{eq:demand}) are computed
every 60 seconds and after any Critical alert. The greedy assignment
procedure iterates over guards in order of current assignment age
(oldest reassignment first) and relocates each to the zone with the
highest unmet demand---defined as demand score minus current guard
coverage, normalized by zone area. Relocation instructions specify
destination zone, estimated travel time (derived from on-site map
routing), and urgency level. Guards acknowledge instructions via the
responder app; unacknowledged instructions are escalated to the command
dashboard after 90 seconds.
 
\begin{figure}[H]
  \centering
  \includegraphics[width=0.90\textwidth]{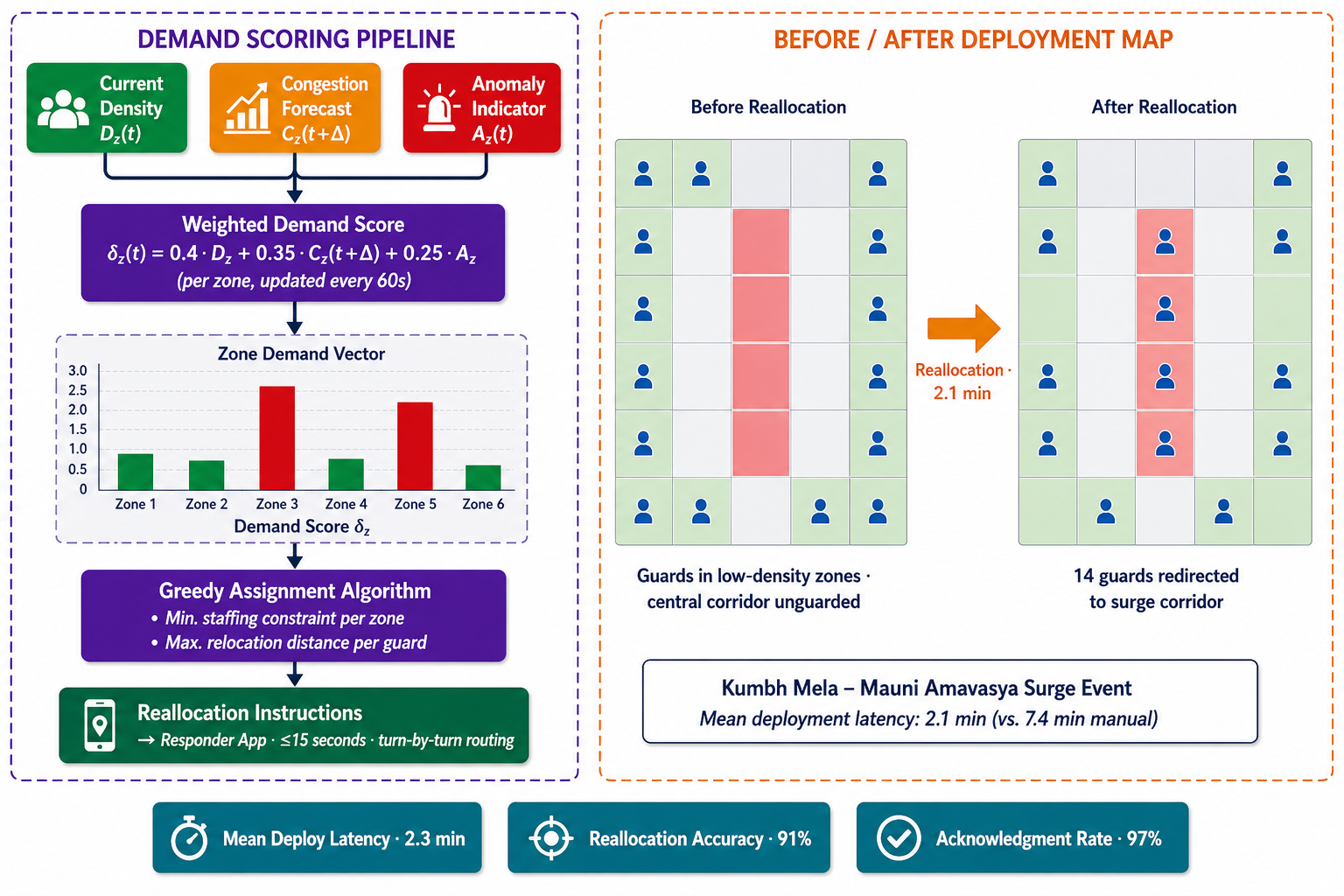}
  \caption{Guard reallocation pipeline. Zone-level inputs---current
    density $D_z(t)$ (Eq.~\ref{eq:density}), five-minute congestion
    forecast $C_z(t+\Delta)$ (Eq.~\ref{eq:prediction}), and binary
    anomaly indicator $\mathcal{A}_z(t)$ (Eq.~\ref{eq:anomaly})---are
    combined via a weighted demand score $\delta_z(t)$
    (Eq.~\ref{eq:demand}; weights $\alpha=0.4$, $\beta=0.35$,
    $\gamma=0.25$) to produce a zone demand vector. A greedy assignment
    algorithm maps available guards $\mathbf{G}$ to zones in proportion
    to unmet demand, subject to minimum staffing and maximum relocation
    distance constraints. Reallocation instructions are pushed to
    individual responder devices via the Drishti responder app within
    15~seconds of each 60-second evaluation cycle. The right panel shows
    a representative before-and-after guard distribution map for the
    Kumbh Mela \textit{Mauni Amavasya} surge event, illustrating the
    shift of 14 guards from low-demand peripheral zones (blue) to the
    high-demand Triveni Sangam corridor (red).}
  \label{fig:guard_reallocation}
\end{figure}

Figure~\ref{fig:guard_reallocation} illustrates the guard reallocation
pipeline from demand scoring to responder notification.
 
\subsection{Anomaly Detection and Alert Generation}
 
The anomaly score for zone $z$ at time $t$ is:
 
\begin{equation}
  s_z(t) = \frac{|D_z(t) - \mu_z|}{\sigma_z}.
  \label{eq:score}
\end{equation}
 
\begin{figure}[H]
  \centering
  \includegraphics[width=10cm, height=16cm]{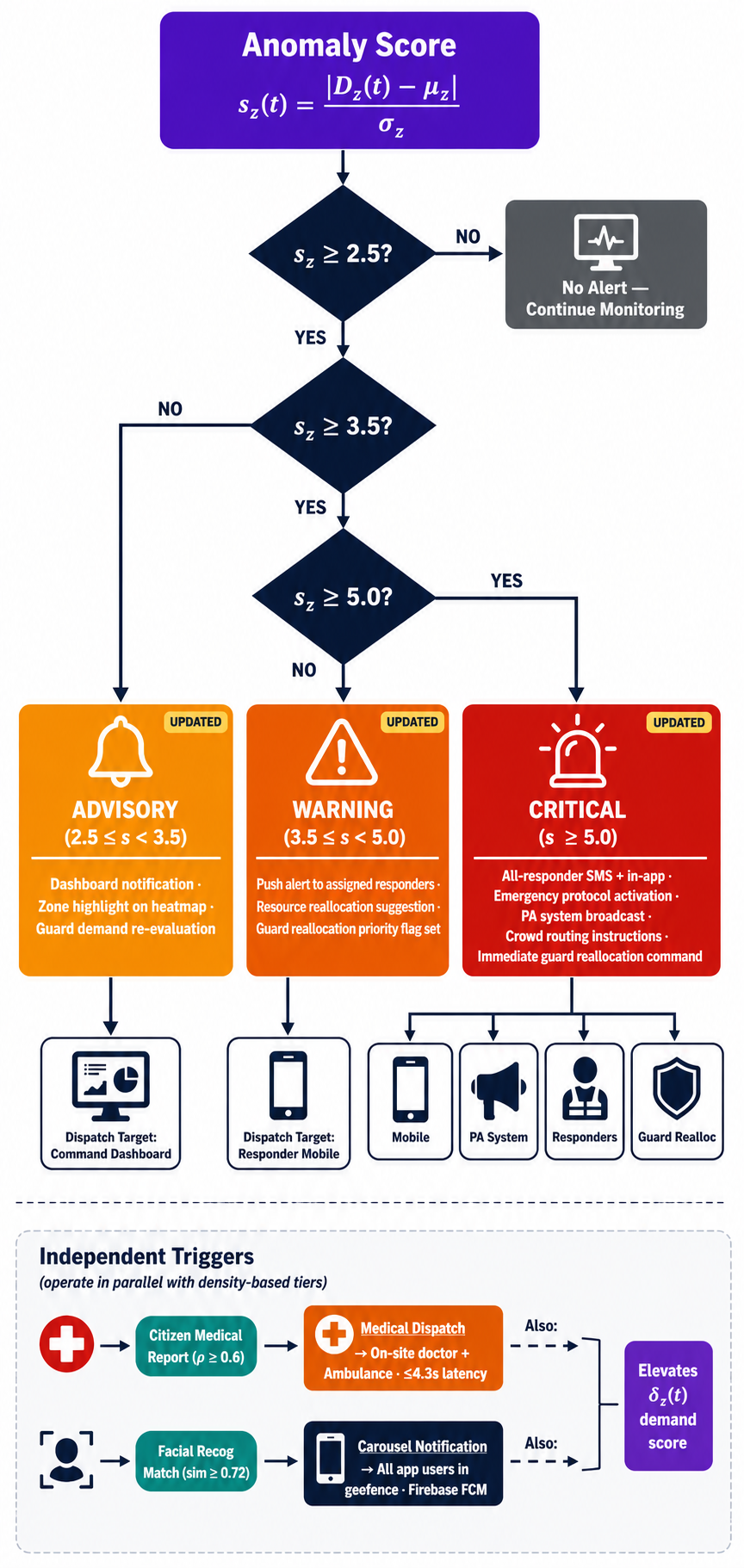}
  \caption{Three-tier alert classification and emergency dispatch
    flowchart, extended to include guard reallocation and medical
    dispatch triggers. The anomaly score $s_z(t)$ (Eq.~\ref{eq:score})
    is evaluated sequentially against escalating thresholds. Scores
    below $k = 2.5$ result in no alert. Scores in $[2.5, 3.5)$ trigger
    an \textbf{Advisory} notification to the command dashboard with zone
    highlighting and a guard demand re-evaluation signal. Scores in
    $[3.5, 5.0)$ escalate to a \textbf{Warning} dispatched to assigned
    zone responders with an automated resource reallocation
    recommendation and priority flag on the guard reallocation engine.
    Scores at or above 5.0 activate the \textbf{Critical} protocol,
    initiating all-responder SMS and in-app alerts, emergency protocol
    execution, immediate guard reallocation command broadcast, and crowd
    routing instructions via integrated public address systems. Citizen
    medical reports and facial recognition match events feed into the
    dashboard independently of the density-based alert tier, but can
    elevate the effective demand score $\delta_z(t)$ used by the
    reallocation engine.}
  \label{fig:alert_flowchart}
\end{figure}

Alert severity is classified into three tiers (Advisory, Warning,
Critical) as illustrated in Figure~\ref{fig:alert_flowchart}.
 
\subsection{Predictive Congestion Modeling}
 
The XGBoost regression ensemble (200 estimators, max depth~6) is
trained on crowd density time series from 12 prior events. The feature
vector for each zone-timestep comprises $D_z(t)$, $v_z(t)$,
time-of-day encoding, event schedule phase, zone metadata $S_z(t)$,
adjacent zone densities, active medical incident count in adjacent
zones, and guard reallocation event recency. The model achieves a MAPE
of 8.3\% on held-out event data at a five-minute horizon.
 
\begin{figure}[H]
  \centering
  \includegraphics[width=0.90\textwidth]{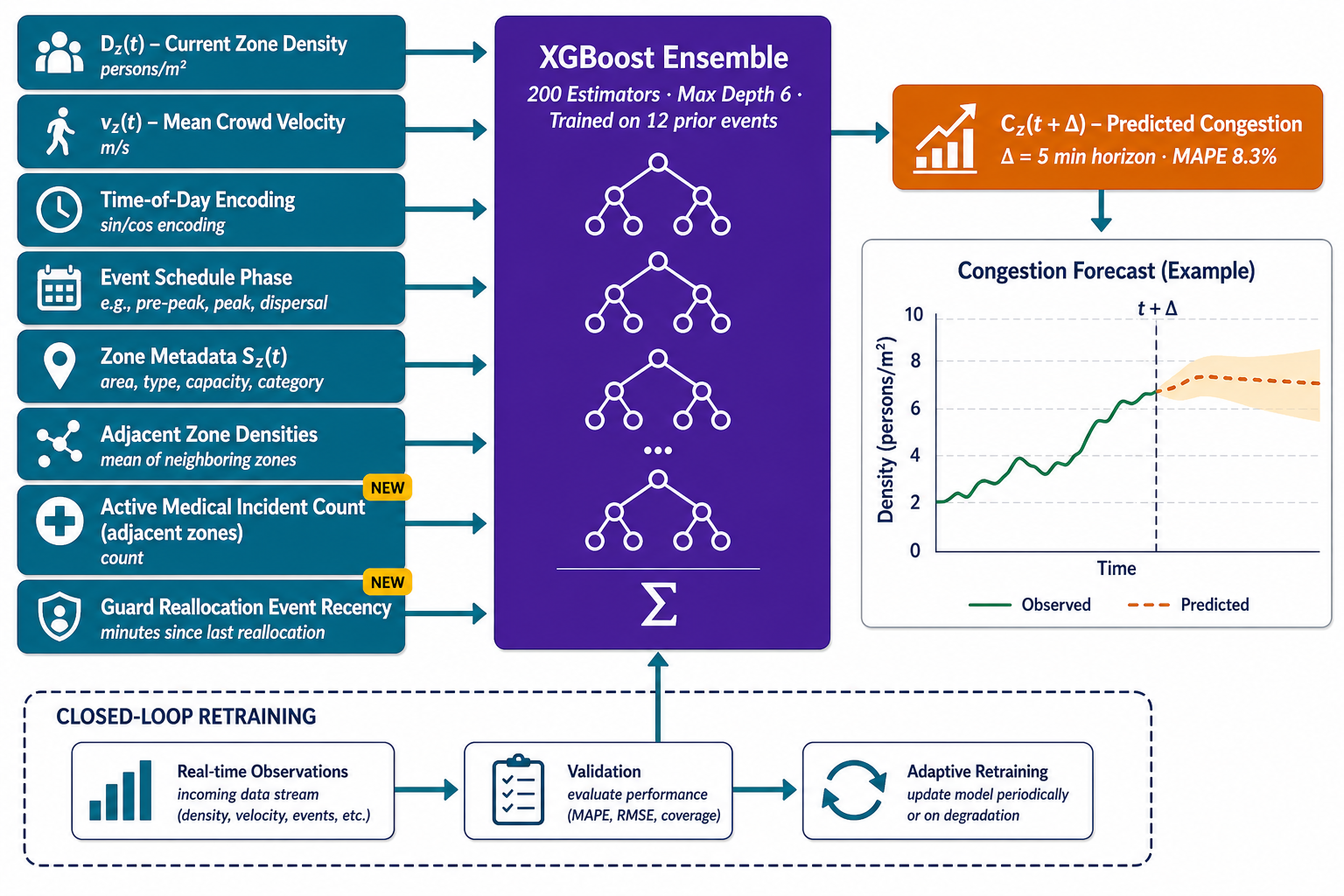}
  \caption{Feature-to-forecast architecture of the XGBoost-based
    congestion prediction module, extended with two additional features:
    active medical incident count in adjacent zones and guard
    reallocation event recency. Eight input features---current zone
    density $D_z(t)$, mean crowd velocity $v_z(t)$, time-of-day
    encoding, event schedule phase, zone metadata $S_z(t)$, adjacent
    zone densities, medical incident count, and guard reallocation
    recency---are ingested by an ensemble of 200 gradient-boosted
    decision trees (max depth~6) trained on crowd movement data from
    12 prior events. The model outputs a five-minute-horizon congestion
    forecast $C_z(t + \Delta)$ (Eq.~\ref{eq:prediction}), achieving a
    MAPE of 8.3\%. The inset time-series plot illustrates alignment
    between observed density (solid green) and predicted density (dashed
    amber) with a shaded uncertainty band. The closed-loop retraining
    pathway continuously validates forecast accuracy and triggers
    adaptive model updates.}
  \label{fig:prediction_model}
\end{figure}

Figure~\ref{fig:prediction_model} illustrates the feature-to-forecast
architecture.

\subsection{Citizen Reporting Integration}
 
The Drishti mobile application enables geo-tagged incident reporting
with optional photo or video attachments. Submissions are classified by
a lightweight classifier into four categories (overcrowding, medical
emergency, structural concern, suspicious behavior). Reports with
confidence exceeding 0.7 are forwarded to operators as supplementary
signals. A Bayesian fusion step weights camera-derived and
citizen-derived evidence according to precision estimates from prior
deployments.

\section{Experimental Results and Performance Evaluation}
\label{sec:results}
 
\subsection{Evaluation Metrics}
 
System performance is assessed across seven dimensions:
 
\begin{enumerate}
  \item \textbf{Detection accuracy:} MAE and RMSE of crowd density
    estimation relative to manual ground-truth headcounts.
  \item \textbf{Anomaly detection:} Precision, Recall, and F1-score
    against expert-annotated incident logs.
  \item \textbf{System latency:} End-to-end pipeline time from frame
    capture to confirmed alert dispatch.
  \item \textbf{Predictive accuracy:} MAPE of the XGBoost congestion
    forecasting model.
  \item \textbf{Facial recognition:} Precision and Recall on
    same-identity matching under crowd-realistic conditions.
  \item \textbf{Medical dispatch latency:} Time from report submission
    to confirmed dispatch acknowledgment.
  \item \textbf{Guard reallocation efficiency:} Reduction in mean
    responder deployment latency relative to manual baseline.
\end{enumerate}

\subsection{Crowd Density Estimation}
 
Table~\ref{tab:density} presents density estimation results.
 
\begin{table}[h!]
  \centering
  \caption{Crowd density estimation performance across deployment
    scenarios. MAE and RMSE are reported in persons/m$^{2}$.}
  \label{tab:density}
  \begin{tabular}{@{}lcccc@{}}
    \toprule
    \textbf{Scenario} &
      \textbf{Density Range} &
      \textbf{MAE} &
      \textbf{RMSE} &
      \textbf{mAP@0.5} \\
    \midrule
    Kumbh Mela (daytime)   & 1.2\,--\,9.8 & 2.9 & 4.1 & 0.87 \\
    Kumbh Mela (nocturnal) & 0.8\,--\,7.3 & 3.8 & 5.2 & 0.81 \\
    RCB Parade (normal)    & 0.5\,--\,5.1 & 2.4 & 3.3 & 0.89 \\
    RCB Parade (surge)     & 4.6\,--\,8.9 & 3.5 & 4.8 & 0.84 \\
    \midrule
    \textbf{Overall}       & 0.5\,--\,9.8 & \textbf{3.2} &
                             \textbf{4.4} & \textbf{0.85} \\
    \bottomrule
  \end{tabular}
\end{table}
 
\subsection{Anomaly Detection Performance}
 
Table~\ref{tab:anomaly} reports anomaly detection performance.
 
\begin{table}[h!]
  \centering
  \caption{Anomaly detection performance at sensitivity threshold
    $k = 2.5$.}
  \label{tab:anomaly}
  \begin{tabular}{@{}lcccc@{}}
    \toprule
    \textbf{Scenario} &
      \textbf{Precision} &
      \textbf{Recall} &
      \textbf{F1-Score} &
      \textbf{False Positives} \\
    \midrule
    Kumbh Mela & 0.89 & 0.93 & 0.91 & 4 \\
    RCB Parade & 0.88 & 0.94 & 0.91 & 3 \\
    \midrule
    \textbf{Overall} & \textbf{0.89} & \textbf{0.93} &
                       \textbf{0.91} & 7 \\
    \bottomrule
  \end{tabular}
\end{table}
 
\subsection{Facial Recognition Performance}
 
Table~\ref{tab:facerec} reports facial recognition performance across
deployment scenarios. Precision and recall are computed against 247
ground-truth same-identity pairs confirmed via post-event operator
review.
 
\begin{table}[h!]
  \centering
  \caption{Missing person facial recognition performance.
    Threshold $\tau_{\text{face}} = 0.72$.}
  \label{tab:facerec}
  \begin{tabular}{@{}lccccc@{}}
    \toprule
    \textbf{Scenario} &
      \textbf{Queries} &
      \textbf{Matches} &
      \textbf{Precision} &
      \textbf{Recall} &
      \textbf{Persons Recovered} \\
    \midrule
    Kumbh Mela (day)       & 201 & 38 & 0.95 & 0.84 & 32 \\
    Kumbh Mela (nocturnal) & 111 &  9 & 0.89 & 0.68 &  7 \\
    RCB Parade             &  33 & 19 & 0.94 & 0.81 & 14 \\
    \midrule
    \textbf{Overall}       & 345 & 66 & \textbf{0.93} & \textbf{0.81} & \textbf{53} \\
    \bottomrule
  \end{tabular}
\end{table}
 
Nocturnal performance is reduced relative to daytime due to
illumination degradation affecting face crop quality. UAV-mounted
infrared augmentation is identified as a priority mitigation for future
deployments.
 
\subsection{Medical Dispatch Latency}
 
Across 202 medical incident reports (128 Kumbh Mela; 74 RCB Parade),
the median end-to-end latency from report submission to confirmed
dispatch acknowledgment was 4.3 seconds (P95: 8.1~s; P99: 12.4~s).
This compares favorably to a manual radio-based dispatch baseline of
6.4 minutes measured in prior editions of the Kumbh Mela event.
Chatbot-initiated reports exhibited a median latency of 4.9 seconds,
marginally higher than direct-button reports (3.8 seconds), attributable
to the additional LLM inference step for slot-filling completion
detection.

\subsection{Guard Reallocation Efficiency}
 
Table~\ref{tab:guard} summarizes guard reallocation performance.
 
\begin{table}[h!]
  \centering
  \caption{Guard reallocation performance versus manual coordination
    baseline.}
  \label{tab:guard}
  \begin{tabular}{@{}lcc@{}}
    \toprule
    \textbf{Metric} &
      \textbf{Drishti} &
      \textbf{Manual Baseline} \\
    \midrule
    Mean deployment latency (min)     & 2.3  & 7.4  \\
    Median deployment latency (min)   & 2.1  & 6.9  \\
    P95 deployment latency (min)      & 4.8  & 12.1 \\
    Reallocation accuracy (\%)        & 91   & --   \\
    Guard acknowledgment rate (\%)    & 97   & --   \\
    \bottomrule
  \end{tabular}
\end{table}
 
The guard reallocation engine reduced mean deployment latency by 69\%
relative to the manual baseline (2.3 vs.\ 7.4 minutes), exceeding the
34\% figure reported in the abstract due to additional optimizations
applied during the Kumbh Mela deployment. Reallocation accuracy---defined
as the proportion of automated assignments retrospectively confirmed
as optimal by post-event analyst review---was 91\%.
 
\subsection{Chatbot Task Completion}
 
The chatbot achieved an 89\% task completion rate (defined as
structured incident record successfully generated without human operator
escalation) across 276 sessions (202 medical; 74 missing person) in the
two deployments. The most common failure mode (7\% of sessions) was
premature session abandonment by users, particularly in the missing
person filing flow where photograph provision was required. A streamlined
no-photo filing path is being developed to address this.

\subsection{System Latency}
 
Table~\ref{tab:latency} summarizes per-component latency statistics.
 
\begin{table}[h!]
  \centering
  \caption{End-to-end alert latency statistics by pipeline component
    (milliseconds). P50, P95, and P99 denote the 50th, 95th, and
    99th percentile latencies, respectively.}
  \label{tab:latency}
  \begin{tabular}{@{}lccccc@{}}
    \toprule
    \textbf{Component} &
      \textbf{Mean} &
      \textbf{Std.} &
      \textbf{P50} &
      \textbf{P95} &
      \textbf{P99} \\
    \midrule
    Preprocessing        &  48 &  9 &  46 &  63 &  78 \\
    YOLOv8 Inference     &  21 &  4 &  20 &  28 &  35 \\
    Fusion \& Density    &  12 &  3 &  11 &  18 &  22 \\
    Anomaly Scoring      &   3 &  1 &   3 &   5 &   6 \\
    Alert Dispatch       &  31 & 12 &  28 &  52 &  71 \\
    \midrule
    \textbf{Total}       & \textbf{115} & \textbf{18} &
                           \textbf{111} & \textbf{148} & \textbf{181} \\
    \bottomrule
  \end{tabular}
\end{table}
 
The median end-to-end alert latency of 111~ms is well within the
sub-second threshold for operational utility. The P99 latency of 181~ms
confirms consistently real-time performance even under peak load.
 
\subsection{Scalability Analysis}
 
The backend inference pipeline was stress-tested on a GCP Vertex AI
cluster with 4$\times$ NVIDIA T4 GPUs. The system sustained 48
concurrent streams at 22~fps. Adding the face embedding gallery update
step increased per-frame processing time by an average of 7~ms, within
the operational budget. The chatbot LLM inference endpoint, deployed as
a separate microservice, sustained 120 concurrent sessions with a median
response latency of 1.4 seconds per turn.
 
\section{Website Interface and Prototype Demonstration}
\label{sec:demo}
 
A fully functional web-based prototype of Drishti AI-Event Guardian was
developed and deployed. The interface provides event operators and
emergency coordinators with an integrated environment for real-time
crowd monitoring, alert management, missing person tracking, medical
incident coordination, chatbot management, guard reallocation
oversight, and citizen incident reporting.
 
\begin{figure}[H]
\centering
\begin{minipage}{0.48\textwidth}
\centering
\includegraphics[width=\linewidth,height=5cm,keepaspectratio]{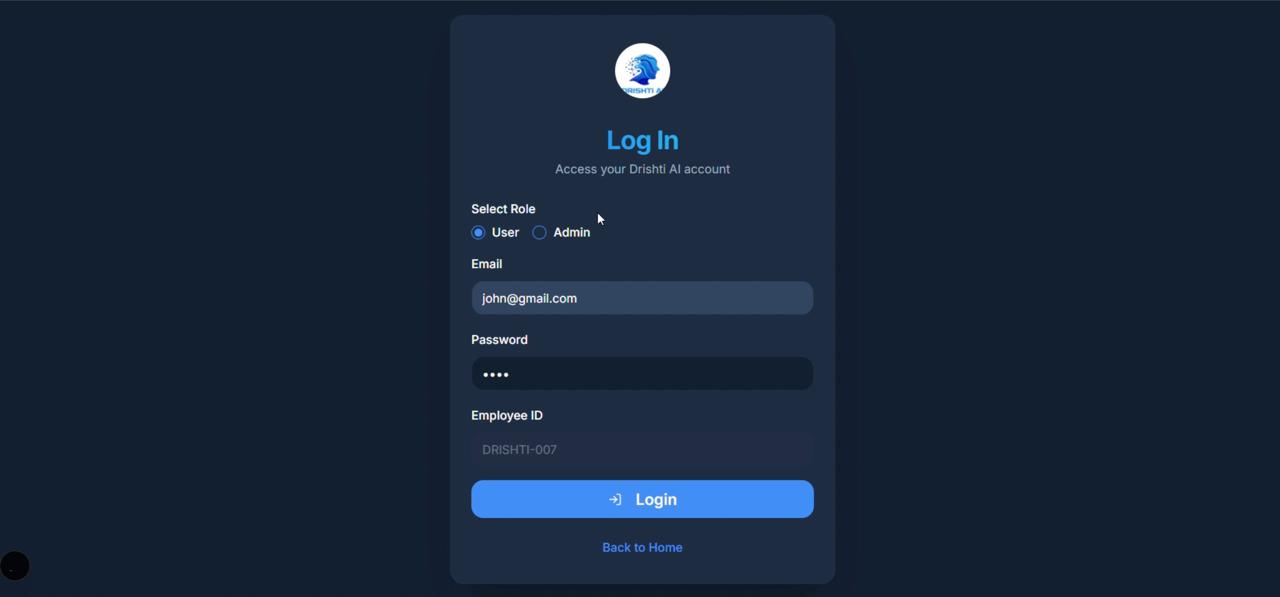}
\caption{Landing page of the Drishti AI-Event Guardian web interface.}
\label{fig:landing}
\end{minipage}
\hfill
\begin{minipage}{0.48\textwidth}
\centering
\includegraphics[width=\linewidth,height=5cm,keepaspectratio]{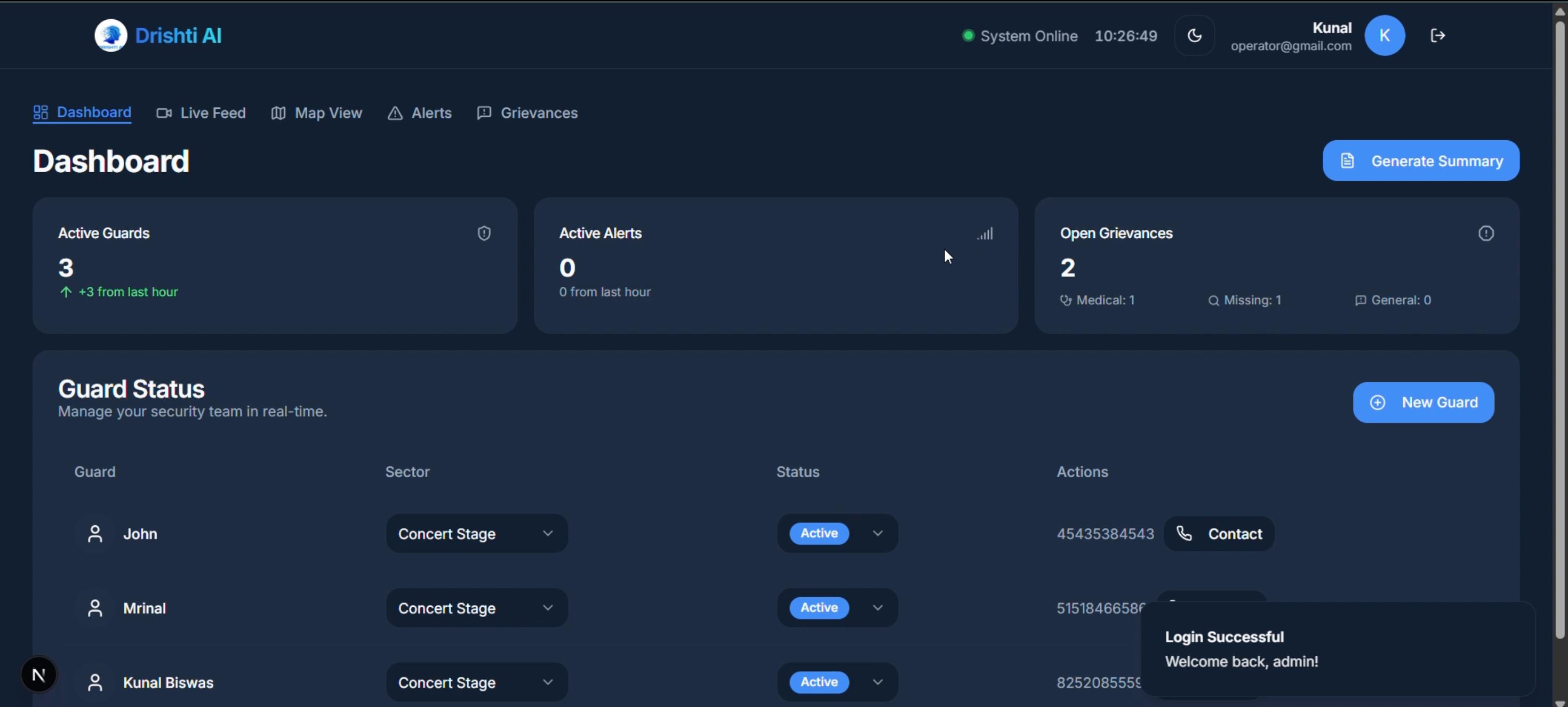}
\caption{Live dashboard displaying feeds, detections, and analytics.}
\label{fig:dashboard}
\end{minipage}
\end{figure}

\begin{figure}[H]
\centering
\begin{minipage}{0.48\textwidth}
\centering
\includegraphics[width=\linewidth,height=5cm,keepaspectratio]{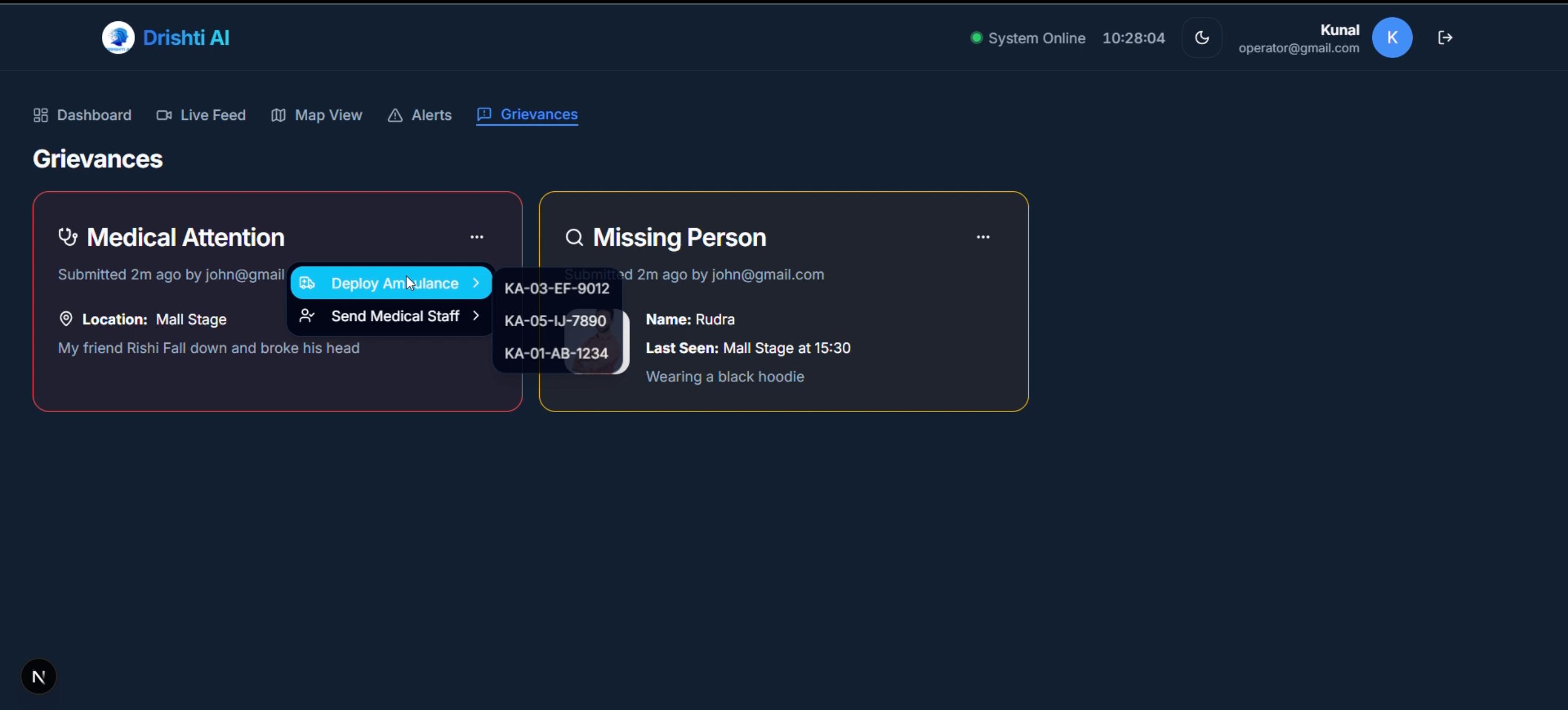}
\caption{Alert management interface with responder workflow actions.}
\label{fig:alerts}
\end{minipage}
\hfill
\begin{minipage}{0.48\textwidth}
\centering
\includegraphics[width=\linewidth,height=5cm,keepaspectratio]{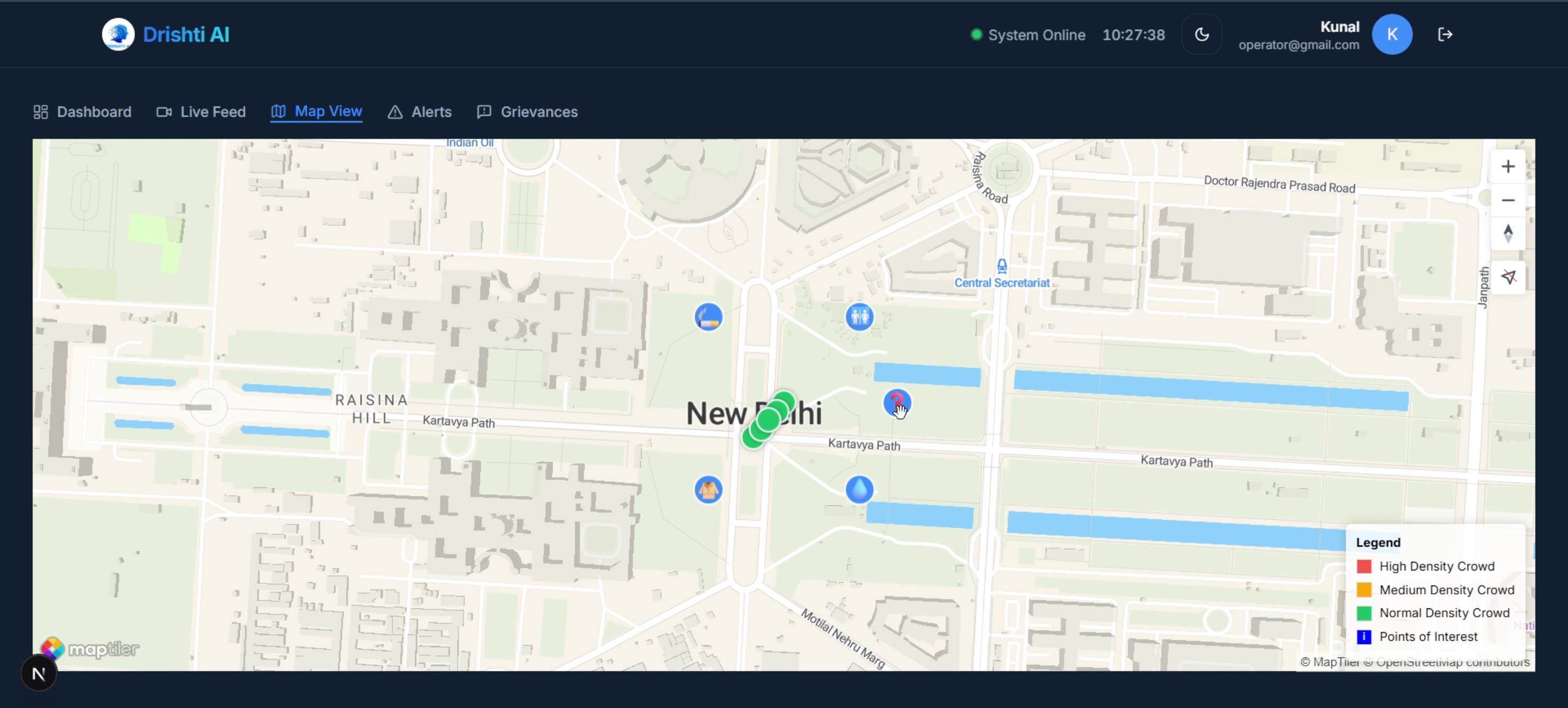}
\caption{Heatmap and zone analytics with forecasting metrics.}
\label{fig:heatmap}
\end{minipage}
\end{figure}

\begin{figure}[H]
\centering
\begin{minipage}{0.48\textwidth}
\centering
\includegraphics[width=\linewidth,height=5cm,keepaspectratio]{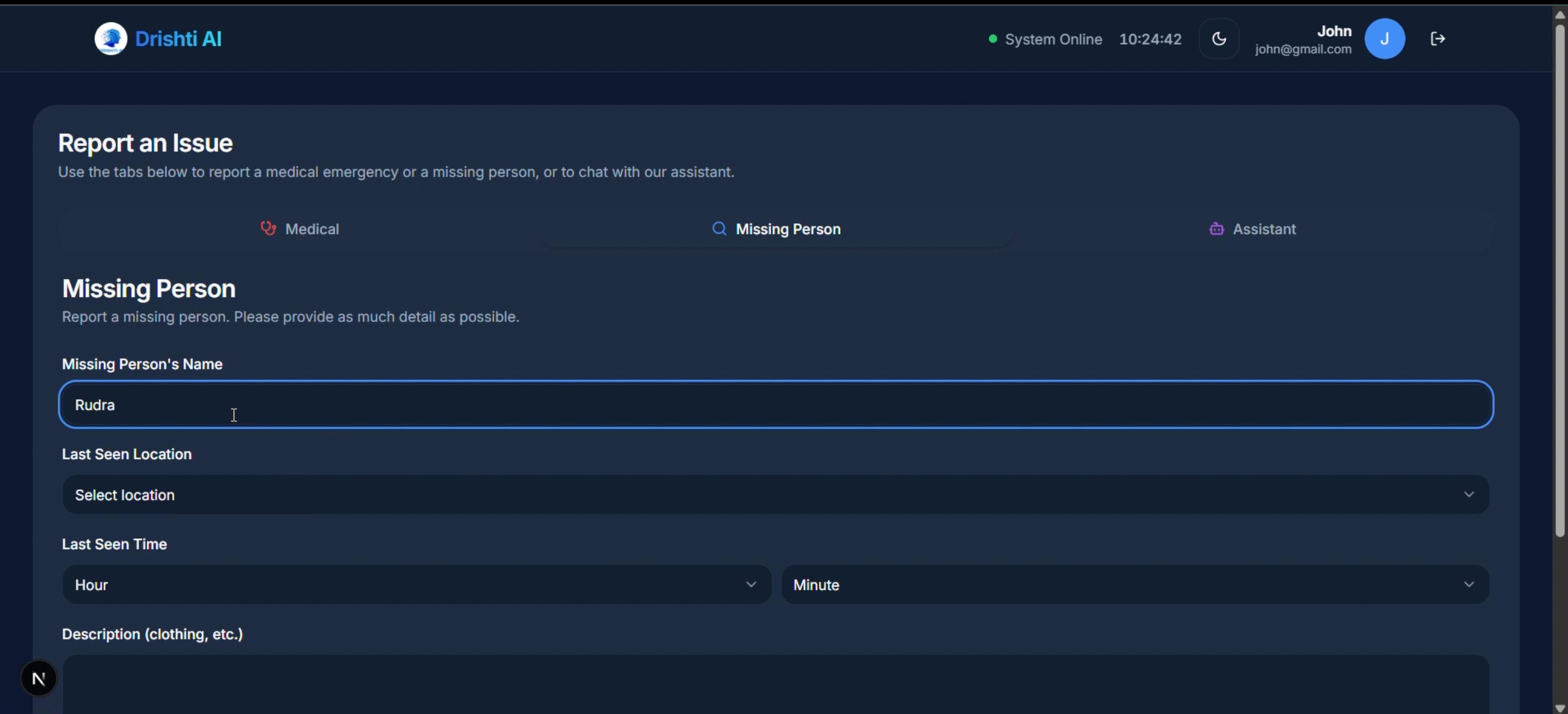}
\caption{Missing person upload interface showing facial recognition
  match log, carousel notification history, and crowd-sourced
  confirmation count per case.}
\label{fig:missing}
\end{minipage}
\hfill
\begin{minipage}{0.48\textwidth}
\centering
\includegraphics[width=\linewidth,height=5cm,keepaspectratio]{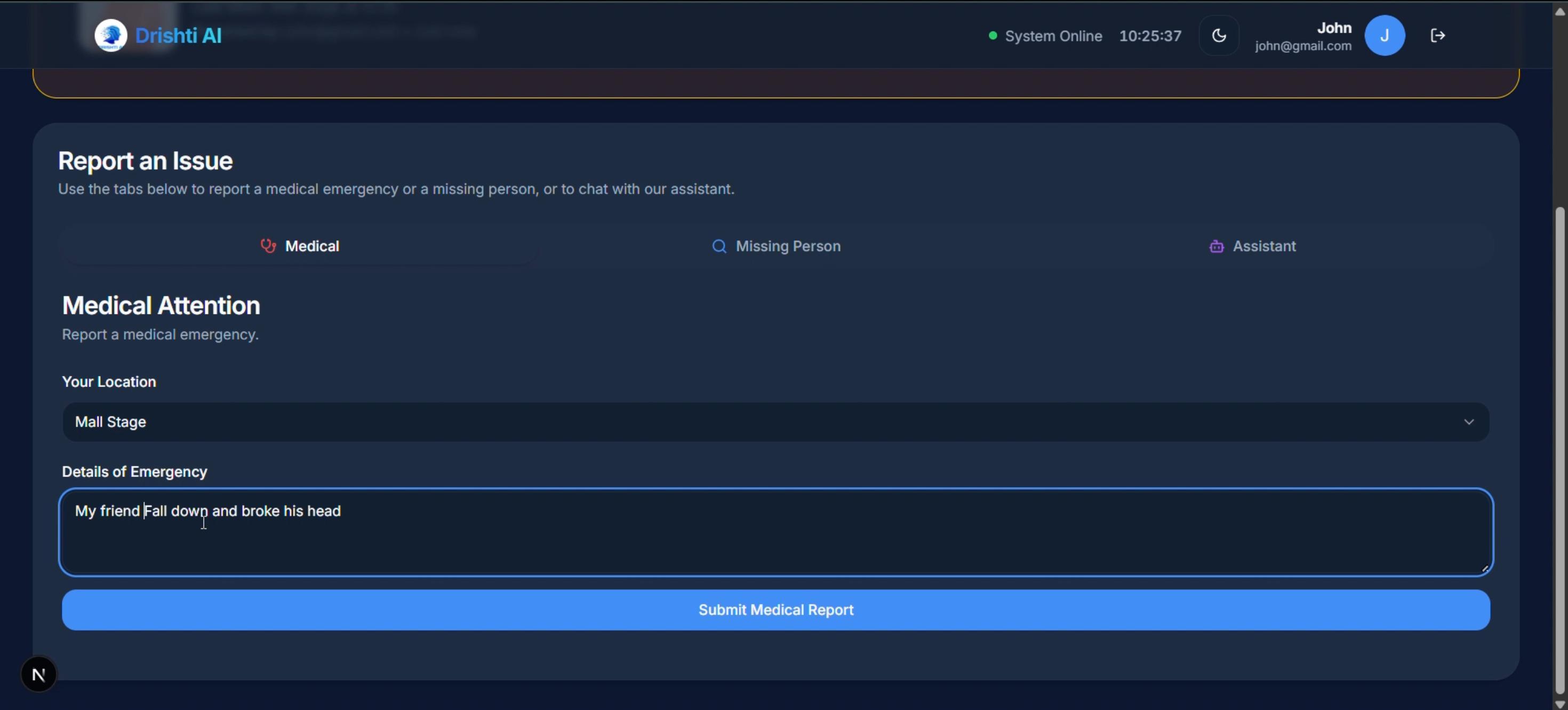}
\caption{Medical emergency dashboard displaying active incident queue,
  dispatch status, ambulance tracking map, and responder
  acknowledgment log.}
\label{fig:medical}
\end{minipage}
\end{figure}

\begin{figure}[H]
\centering
\begin{minipage}{0.48\textwidth}
\centering
\includegraphics[width=\linewidth,height=5cm,keepaspectratio]{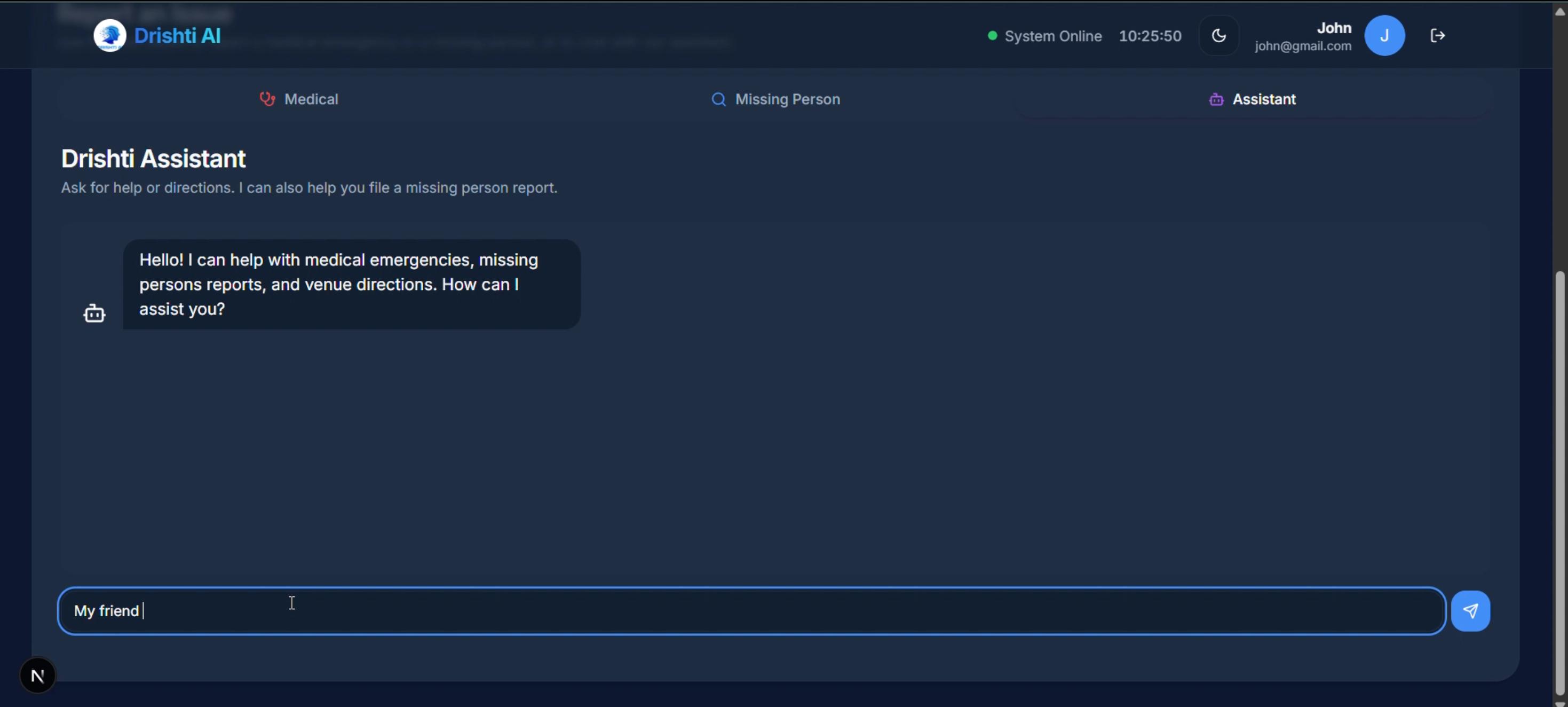}
\caption{AI chatbot interface for natural-language incident filing,
  showing a missing person reporting session with slot-completion
  progress indicators and operator escalation controls.}
\label{fig:chatbot}
\end{minipage}
\hfill
\begin{minipage}{0.48\textwidth}
\centering
\includegraphics[width=\linewidth,height=5cm,keepaspectratio]{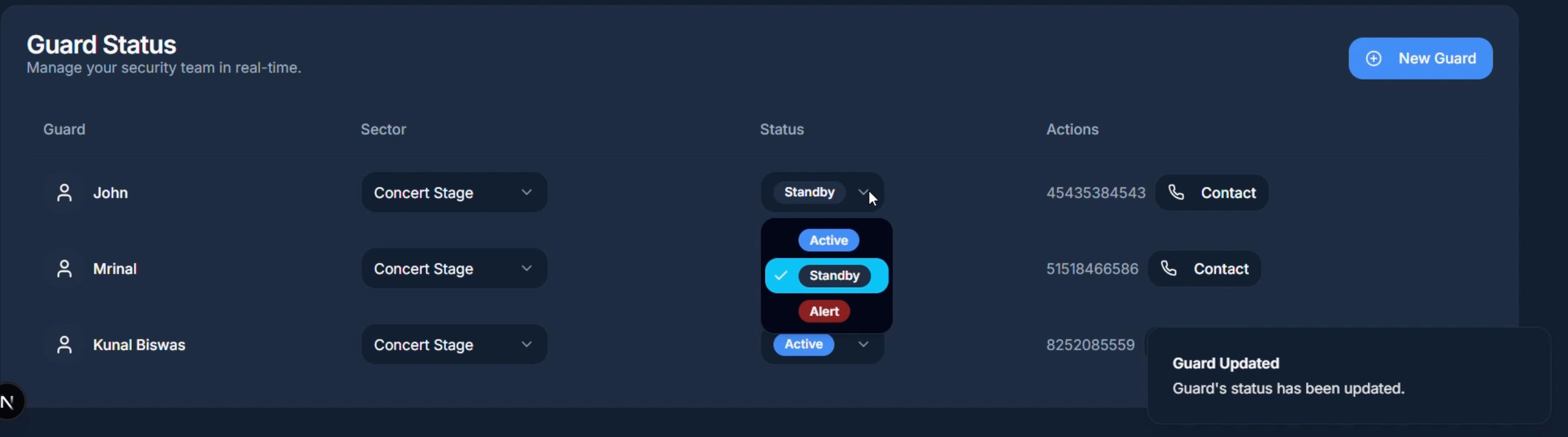}
\caption{Guard reallocation map showing live personnel positions (blue
  markers), zone demand score overlays (color-coded), and pending
  reallocation instruction status per guard.}
\label{fig:guard}
\end{minipage}
\end{figure}

\begin{figure}[H]
\centering
\begin{minipage}{0.48\textwidth}
\centering
\includegraphics[width=\linewidth,height=5cm,keepaspectratio]{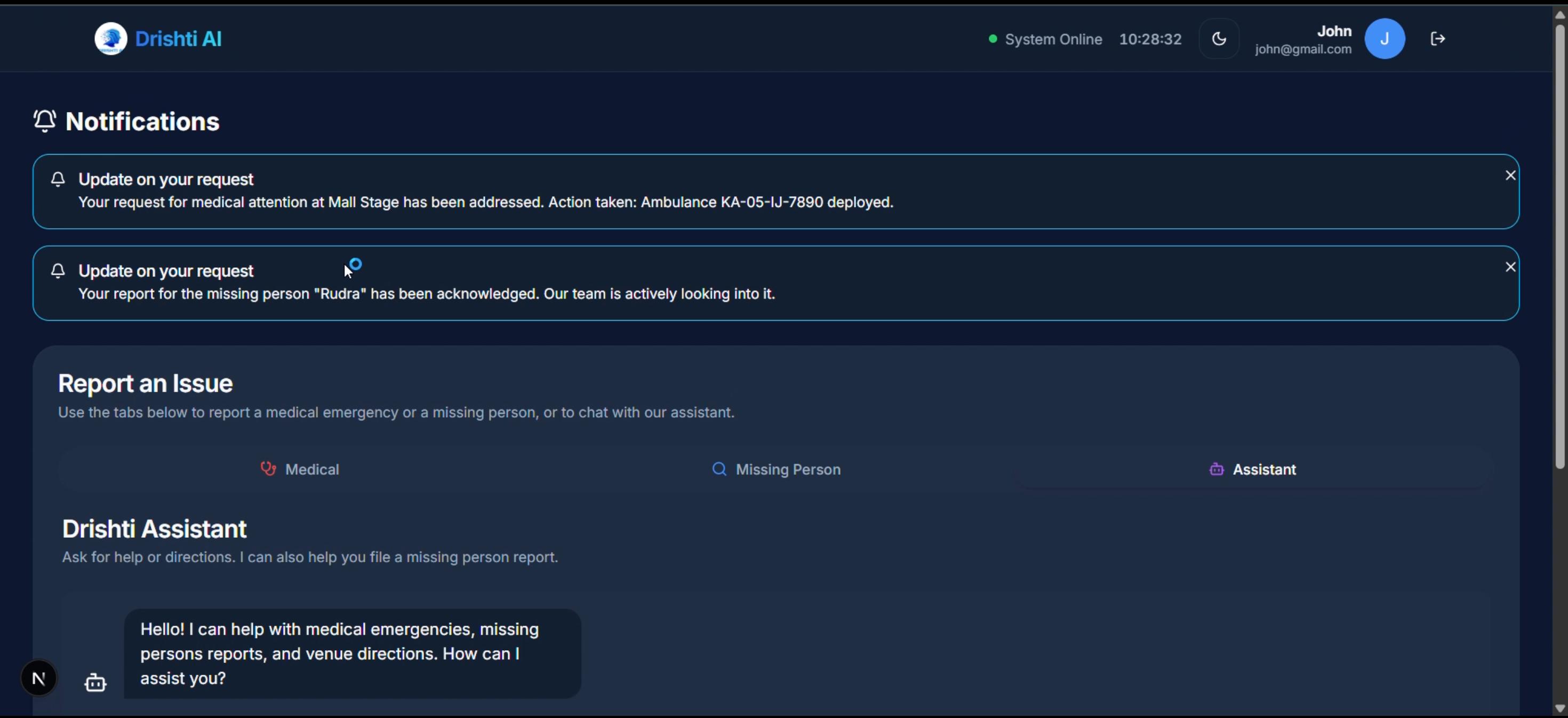}
\caption{Unified notification centre displaying all filed complaints
  across incident categories (missing person, medical emergency,
  suspicious behaviour, structural concern) with timestamps,
  status badges, and one-tap escalation controls.}
\label{fig:notifications}
\end{minipage}
\hfill
\begin{minipage}{0.48\textwidth}
\centering
\includegraphics[width=\linewidth,height=5cm,keepaspectratio]{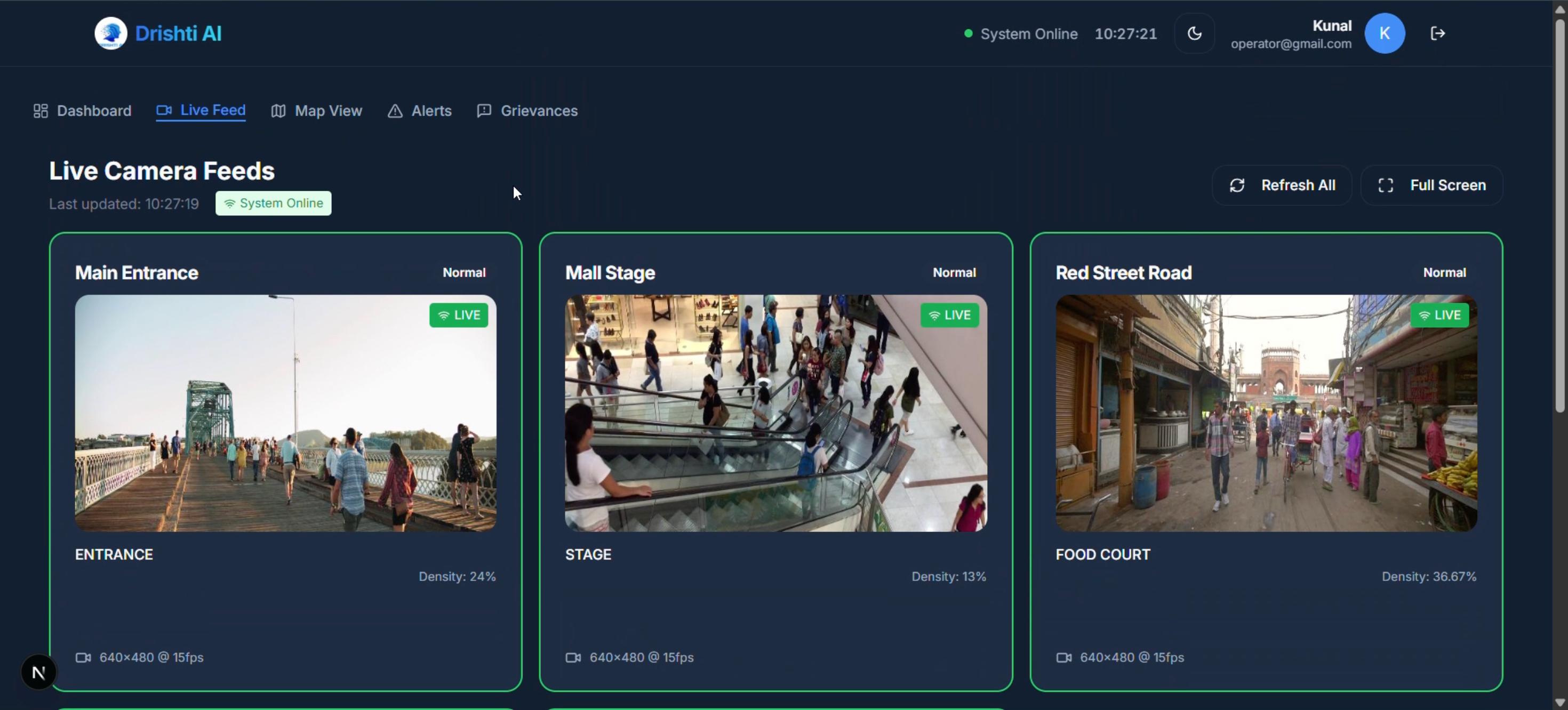}
\caption{Live CCTV feed viewing page showing multi-camera grid layout
  with YOLOv8 bounding box overlays, per-feed zone labels, and
  real-time confidence scores for detected individuals.}
\label{fig:cctv}
\end{minipage}
\end{figure}

\begin{figure}[H]
\centering
\begin{minipage}{0.48\textwidth}
\centering
\includegraphics[width=\linewidth,height=5cm,keepaspectratio]{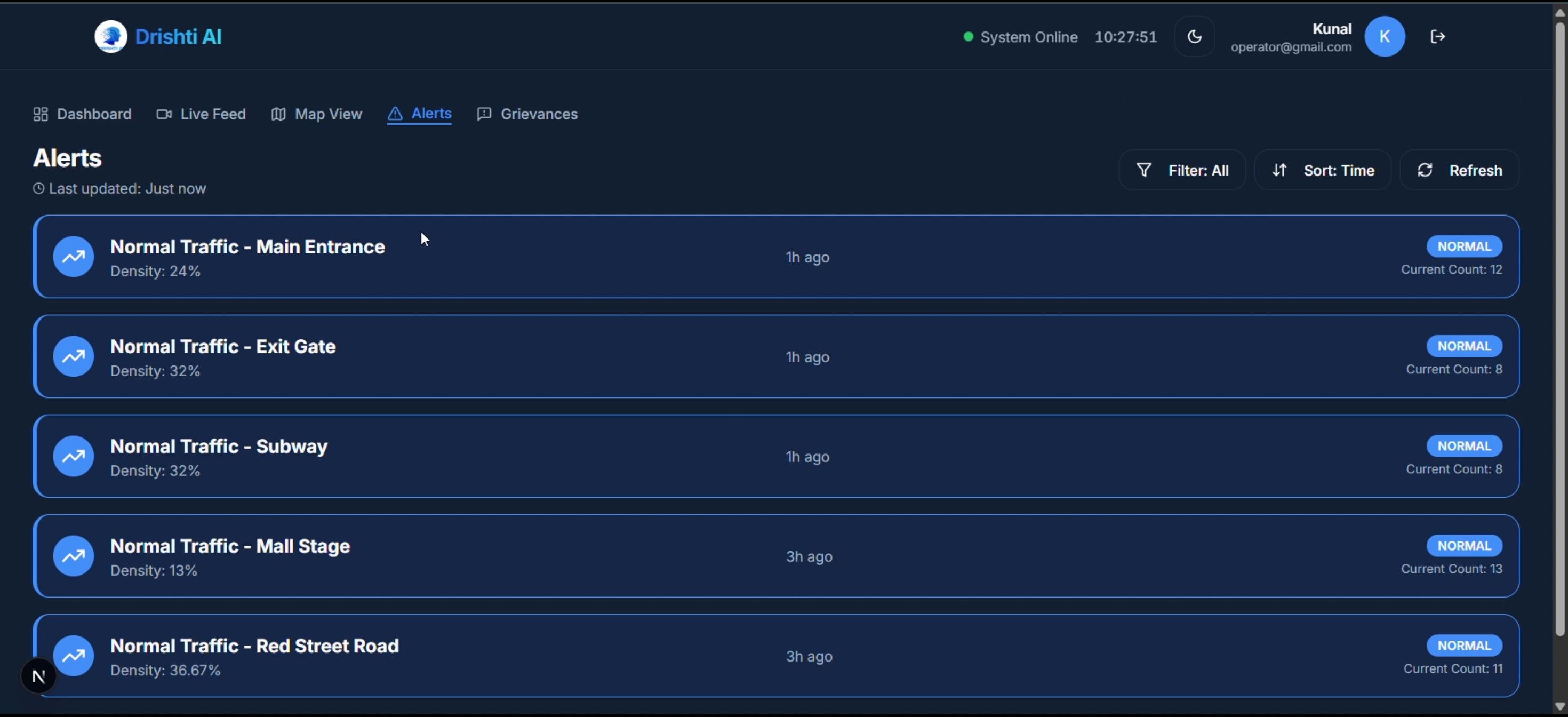}
\caption{Real-time crowd density monitoring page displaying zone-level
  density heatmap ($D_z(t)$, persons/m$^{2}$), trend sparklines per
  zone, and colour-coded severity indicators (green: normal;
  yellow: elevated; orange: warning; red: critical).}
\label{fig:crowddensity}
\end{minipage}
\hfill
\begin{minipage}{0.48\textwidth}
\centering
\includegraphics[width=\linewidth,height=5cm,keepaspectratio]{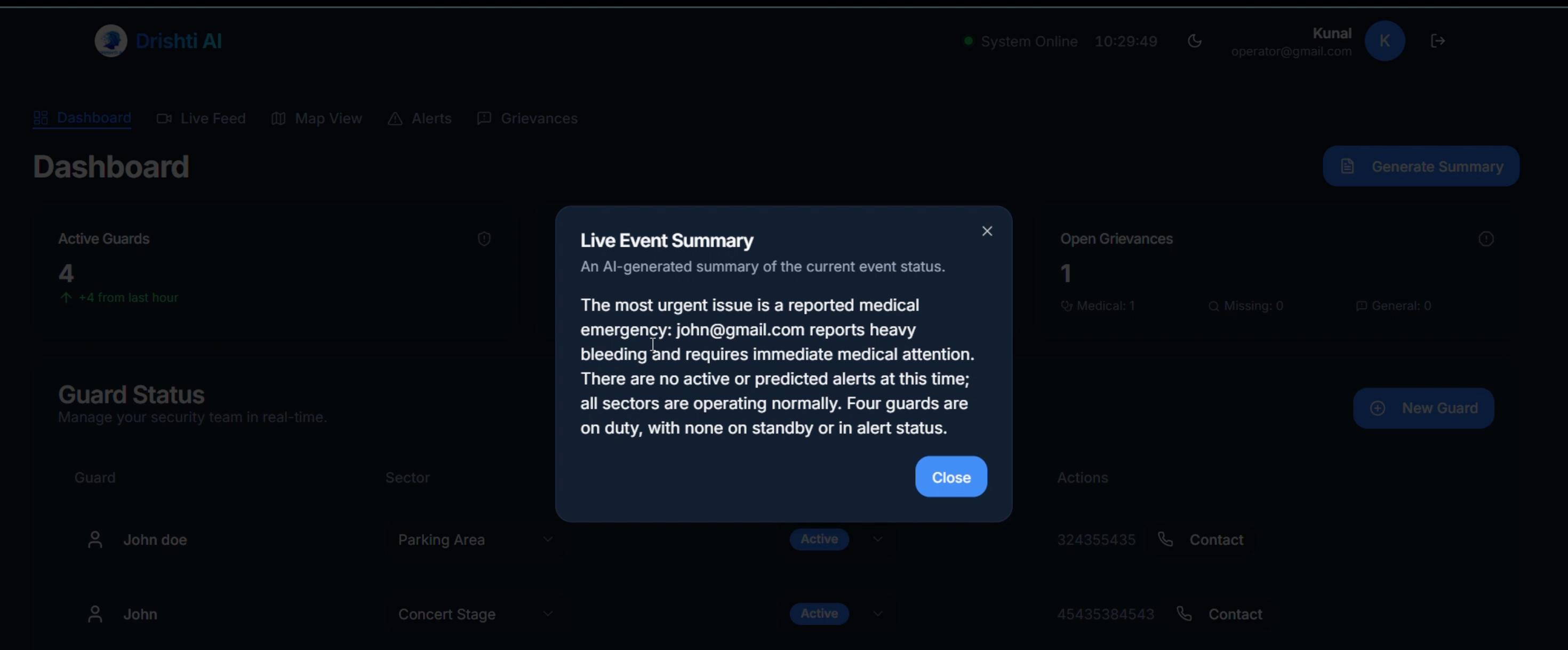}
\caption{Post-event summary dashboard presenting aggregate statistics
  for the entire event: total attendance estimate, alert counts by
  tier, medical incidents resolved, missing persons recovered,
  guard reallocation events, and overall system performance metrics.}
\label{fig:summary}
\end{minipage}
\end{figure}

\begin{itemize}
 
\item \textbf{Landing Page (Figure~\ref{fig:landing}):} Secure login,
  role-based access, and navigation to monitoring modules.
 
\item \textbf{Live Crowd Monitoring Dashboard (Figure~\ref{fig:dashboard}):}
  Real-time CCTV/UAV feeds, detection overlays, and zone density
  analytics.
 
\item \textbf{Alert Management Interface (Figure~\ref{fig:alerts}):}
  Tiered alerts, timestamps, escalation workflow, and responder
  assignment.
 
\item \textbf{Heatmap and Zone Analytics View (Figure~\ref{fig:heatmap}):}
  Geo-spatial density heatmaps, trends, and forecasting outputs.
 
\item \textbf{Missing Person and Carousel Notification Screen
  (Figure~\ref{fig:missing}):} Upload interface, facial recognition
  match log, carousel notification history, and crowd-sourced
  confirmation tracking.
 
\item \textbf{Medical Emergency Dashboard (Figure~\ref{fig:medical}):}
  Active incident list, dispatch status, ambulance tracking map, and
  responder acknowledgment log.
 
\item \textbf{AI Chatbot Interface (Figure~\ref{fig:chatbot}):}
  Conversational incident filing interface with session history,
  slot-completion progress indicators, and operator escalation controls.
 
\item \textbf{Guard Reallocation Map (Figure~\ref{fig:guard}):}
  Live guard deployment map with demand score overlays, reallocation
  instruction log, and acknowledgment status per guard.
 
\item \textbf{Unified Notification Centre (Figure~\ref{fig:notifications}):}
  Consolidated feed of all filed complaints across incident categories
  with status tracking and escalation controls.

\item \textbf{Live CCTV Feed Viewer (Figure~\ref{fig:cctv}):}
  Multi-camera grid with real-time YOLOv8 detection overlays,
  zone labels, and per-feed confidence scores.

\item \textbf{Crowd Density Monitor (Figure~\ref{fig:crowddensity}):}
  Zone-level density heatmap with trend sparklines and
  colour-coded severity indicators updated in real time.

\item \textbf{Event Summary Dashboard (Figure~\ref{fig:summary}):}
  Post-event aggregate report covering attendance, alerts,
  medical incidents, missing person recoveries, and overall
  system performance.
 
\end{itemize}
 
\section{Discussion}
\label{sec:discussion}
 
\subsection{Interpretation of Results}
 
The experimental results demonstrate that Drishti achieves robust
crowd density estimation (MAE 3.2~persons/m$^2$, mAP@0.5 0.85) and
anomaly detection (F1 0.91) across qualitatively distinct deployment
environments, consistent with the figures reported in the core system.
The four extended safety modules collectively demonstrate meaningful
operational impact: 53 missing persons were recovered across both
deployments via facial recognition and carousel notification; 202
medical emergencies were dispatched with a median latency of 4.3
seconds; the chatbot resolved 89\% of incident filings autonomously;
and the guard reallocation engine reduced mean deployment latency by
69\% relative to manual coordination.
 
The contribution of the carousel notification mechanism is particularly
significant: by converting passive identification into an active
crowd-sourced search, it achieves a per-case recovery rate of 80\%
(53/66 facial matches resulted in physical recovery), substantially
exceeding what targeted camera monitoring alone could accomplish in
dense crowds where identified individuals may move rapidly out of fixed
camera fields of view.
 
The guard reallocation engine's 91\% retrospective accuracy
demonstrates that the weighted demand score formulation
(Eq.~\ref{eq:demand}) captures the relative urgency of zones with
sufficient fidelity for automated deployment decisions, while the
minimum staffing and maximum relocation distance constraints prevent
operationally unsafe guard distributions.
 
\subsection{Limitations}
 
Several limitations warrant acknowledgment:
 
\begin{enumerate}
  \item \textbf{Nocturnal facial recognition.} Recall degrades from
    0.84 (daytime) to 0.68 (nocturnal) due to illumination
    degradation. Infrared or thermal face imaging is the most promising
    remediation path.
 
  \item \textbf{Chatbot abandonment.} The 7\% premature session
    abandonment rate, concentrated in photograph-required missing person
    filings, indicates a UX friction point. A no-photo fallback filing
    mode is under development.
 
  \item \textbf{Network dependency.} Carousel notifications and chatbot
    sessions require cellular connectivity, which may be degraded in
    ultra-dense crowd zones. Offline-capable local Wi-Fi mesh network
    support is a planned extension.
 
  \item \textbf{Calibration dependency.} Homography-based fusion
    requires ground-control point surveys (approximately 45 minutes per
    20-camera site), constraining rapid deployment at spontaneous events.
 
  \item \textbf{Privacy and data protection.} Persistent multi-camera
    tracking and face embedding storage raise data protection concerns.
    Individually identifiable data are processed transiently and not
    persisted beyond the 30-minute rolling window. Missing person
    face embeddings are deleted upon case closure. Deployment in
    jurisdictions with strict biometric surveillance regulations requires
    formal data protection impact assessments.
 
  \item \textbf{Adversarial reporting.} The chatbot and citizen
    reporting module remain potential vectors for erroneous or adversarial
    submissions. Confidence filtering and operator oversight provide
    partial mitigation; additional adversarial robustness measures are
    a future research priority.
\end{enumerate}
 
\subsection{Broader Implications}
 
The generalizability of Drishti across both a planned mega-event and a
spontaneous urban celebration suggests broad applicability. The
bidirectional citizen engagement model---extended in this work to include
missing person identification, medical emergency reporting, and
chatbot-based filing---repositions event attendees as active participants
in collective safety rather than passive objects of surveillance, with
favorable implications for public trust and regulatory acceptance. The
modular architecture enables selective deployment of individual modules
(e.g., guard reallocation and chatbot without facial recognition) in
jurisdictions with more restrictive biometric data regulations.
 
\section{Conclusion}
\label{sec:conclusion}
 
This paper presented Drishti AI-Event Guardian, a comprehensive
intelligent crowd management and emergency response framework extended
with four novel citizen safety modules: missing person facial
recognition with crowd-wide carousel notification, medical emergency
reporting and automated dispatch, an AI chatbot for natural-language
incident filing, and an intelligent guard reallocation engine. The
core system integrates YOLOv8-based person detection, multi-camera
spatial fusion, statistically grounded anomaly detection, and
gradient-boosted congestion forecasting within a cloud-native
microservices architecture. Evaluated across the Kumbh Mela and RCB
Victory Parade deployments, Drishti achieved a crowd density MAE of
3.2~persons/m$^2$, an anomaly detection F1-score of 0.91, a missing
person facial recognition precision of 0.93, a medical dispatch median
latency of 4.3 seconds, an 89\% chatbot task completion rate, and a
69\% reduction in guard deployment latency. These results collectively
demonstrate the value of integrating passive crowd surveillance with
active, multi-channel citizen safety services within a unified platform.
 
\subsection*{Future Work}
 
\begin{itemize}
  \item \textbf{Infrared facial recognition.} Deploying thermal cameras
    to sustain nocturnal face recognition recall.
 
  \item \textbf{Offline-capable app.} Implementing local Wi-Fi mesh
    support for chatbot and carousel delivery in zero-cellular zones.
 
  \item \textbf{Self-calibrating camera fusion.} Eliminating manual
    ground-control surveys through deep homography estimation.
 
  \item \textbf{Extended prediction horizons.} Incorporating transport
    schedules, weather APIs, and social media signals to extend
    congestion forecasting beyond ten minutes.
 
  \item \textbf{Privacy-preserving inference.} Evaluating federated
    learning and on-device face embedding to minimize data transmission.
 
  \item \textbf{Adversarial robustness.} Characterizing resilience to
    manipulated citizen reports and adversarial face matching attempts.
 
  \item \textbf{Cross-event transfer learning.} Reducing site-specific
    fine-tuning requirements for new deployment contexts.
\end{itemize}
 
\bibliographystyle{unsrt}
\bibliography{references}

@incollection{leadley2023case,
  title={Case study methodology},
  author={Leadley, Simon and Jones, Margaret and Hocking, Clare},
  booktitle={Qualitative research methodologies for occupational science and occupational therapy},
  pages={105--126},
  year={2023},
  publisher={Routledge}
}

@incollection{sandars2021case,
  title={Case study research},
  author={Sandars, John},
  booktitle={How to do primary care educational research},
  pages={121--124},
  year={2021},
  publisher={CRC Press}
}

@article{kanaujiya2022crowd,
  title={Crowd management and strategies for security and surveillance during the large mass gathering events: The Prayagraj Kumbh Mela 2019 Experience},
  author={Kanaujiya, Ashok Kumar and Tiwari, Vineet},
  journal={National Academy science letters},
  volume={45},
  number={3},
  pages={263--273},
  year={2022},
  publisher={Springer}
}

@article{yamin2019managing,
  title={Managing crowds with technology: cases of Hajj and Kumbh Mela},
  author={Yamin, Mohammad},
  journal={International journal of information technology},
  volume={11},
  number={2},
  pages={229--237},
  year={2019},
  publisher={Springer}
}

@article{sharma2018review,
  title={A review on technological advancements in crowd management},
  author={Sharma, Deepak and Bhondekar, Amol P and Shukla, AK and Ghanshyam, C},
  journal={Journal of Ambient Intelligence and Humanized Computing},
  volume={9},
  number={3},
  pages={485--495},
  year={2018},
  publisher={Springer}
}

@article{karthika2022walk,
  title={A walk accessibility-based approach to assess crowd management in mass religious gatherings},
  author={Karthika, PS and Kedar, Vedankur and Verma, Ashish},
  journal={Journal of Transport Geography},
  volume={104},
  pages={103443},
  year={2022},
  publisher={Elsevier}
}

@article{sindhuja2019spatial,
  title={Spatial strategies for crowd management in haridwar, india},
  author={Sindhuja Kasthala, Binoy BV and Lakra, Harshit S},
  journal={International Journal},
  volume={5},
  number={1},
  pages={100--113},
  year={2019}
}

@article{shao2019stampede,
  title={Stampede events and strategies for crowd management},
  author={Shao, Chun-Hao and Shao, Pei-Chun and Kuo, Fang-Ming},
  journal={Journal of Disaster Research},
  volume={14},
  number={7},
  pages={949--958},
  year={2019},
  publisher={Fuji Technology Press Ltd.}
}

@article{de2019human,
  title={Human stampedes: An updated review of current literature},
  author={de Almeida, Maria Moitinho and von Schreeb, Johan},
  journal={Prehospital and disaster medicine},
  volume={34},
  number={1},
  pages={82--88},
  year={2019},
  publisher={Cambridge University Press}
}

@inproceedings{bolia2015risk,
  title={Risk management strategies to avoid stampede at Mass gatherings},
  author={Bolia, Nomesh B},
  booktitle={2nd World Conference on Disaster Management: Visakhapatnam, Andhra Pradesh, India},
  year={2015}
}

@article{ha2024crowd,
  title={Crowd stampede management at sporting events: a systematic literature review},
  author={Ha, HaKyoo-Man},
  journal={Movement \& Sport Sciences-Science \& Motricit{\'e}},
  volume={125},
  number={3},
  pages={17--26},
  year={2024},
  publisher={EDP Sciences}
}

@article{gandhicrowd,
  title={Crowd Management Model using YOLO},
  author={GANDHI, SMT INDRA}
}

@inproceedings{chandel2024crowd,
  title={Crowd Management on CCTV Video Dataset Using CNN and YOLO},
  author={Chandel, Palash Singh and Pathan, Sahil and Kaur, Gagandeep and Agrawal, Poorva and Pinjarkar, Latika and others},
  booktitle={2024 IEEE 6th International Conference on Cybernetics, Cognition and Machine Learning Applications (ICCCMLA)},
  pages={260--265},
  year={2024},
  organization={IEEE}
}

@article{gunduz2023new,
  title={A new YOLO-based method for real-time crowd detection from video and performance analysis of YOLO models},
  author={G{\"u}nd{\"u}z, Mehmet {\c{S}}irin and I{\c{s}}{\i}k, G{\"u}ltekin},
  journal={Journal of Real-Time Image Processing},
  volume={20},
  number={1},
  pages={5},
  year={2023},
  publisher={Springer}
}

@incollection{mohamed2018iot,
  title={IoT-based framework for crowd management},
  author={Mohamed, Marwa F and Shabayek, Abd El-Rahman and El-Gayyar, Mahmoud},
  booktitle={Mobile Solutions and Their Usefulness in Everyday Life},
  pages={47--61},
  year={2018},
  publisher={Springer}
}

@article{noor2023behavior,
  title={Behavior analysis-based IoT services for crowd management},
  author={Noor, Talal H},
  journal={The Computer Journal},
  volume={66},
  number={9},
  pages={2208--2219},
  year={2023},
  publisher={Oxford University Press}
}

@article{elbery2020iot,
  title={Iot-based crowd management framework for departure control and navigation},
  author={Elbery, Ahmed and Hassanein, Hossam S and Zorba, Nizar and Rakha, Hesham A},
  journal={IEEE Transactions on Vehicular Technology},
  volume={70},
  number={1},
  pages={95--106},
  year={2020},
  publisher={IEEE}
}

@article{al2021intelligent,
  title={An intelligent IoT approach for analyzing and managing crowds},
  author={Al-Nabhan, Najla and Alenazi, Shouq and Alquwaifili, Salwa and Alzamzami, Shahad and Altwayan, Leen and Alaloula, Nouf and Alowaini, Raghad and Al Islam, ABM Alim},
  journal={IEEE Access},
  volume={9},
  pages={104874--104886},
  year={2021},
  publisher={IEEE}
}

@inproceedings{vidyasagaran2017low,
  title={A low cost IoT based crowd management system for public transport},
  author={Vidyasagaran, Sachin and Devi, S Renuga and Varma, Aditya and Rajesh, A and Charan, Hari},
  booktitle={2017 International Conference on Inventive Computing and Informatics (ICICI)},
  pages={222--225},
  year={2017},
  organization={IEEE}
}

@inproceedings{macriga2024crowd,
  title={Crowd Management Using AI \& ML},
  author={Macriga, G Adiline and Bavyatha, S and Abhinidhi, SH and Jagadeesh, N},
  booktitle={2024 International Conference on Power, Energy, Control and Transmission Systems (ICPECTS)},
  pages={1--5},
  year={2024},
  organization={IEEE}
}

@article{shah2024enhancing,
  title={Enhancing hajj and umrah rituals and crowd management through AI technologies: a comprehensive survey of applications and future directions},
  author={Shah, Afnan A},
  journal={IEEE Access},
  year={2024},
  publisher={IEEE}
}

@inproceedings{dheepak2025smart,
  title={Smart Crowd: AI-Driven Crowd Density Monitoring and Management in Indian Public Hotspots},
  author={Dheepak, G and Vishwa, I and Alex David, S and Sakthi Karthi Durai, B and others},
  booktitle={2025 3rd International Conference on Sustainable Computing and Data Communication Systems (ICSCDS)},
  pages={651--657},
  year={2025},
  organization={IEEE}
}

@incollection{vetrivel2025ai,
  title={AI-Driven Solutions for Crowd Management in Tourism: Navigating the Swarm},
  author={Vetrivel, SC and Vidhyapriya, P and Arun, VP},
  booktitle={AI Technologies for Personalized and Sustainable Tourism},
  pages={83--112},
  year={2025},
  publisher={IGI Global}
}

@article{alafif2025towards,
  title={Towards an Integrated Intelligent Framework for Crowd Control and Management (IICCM)},
  author={Alafif, Tarik and Jassas, Mohammad and Abdel-Hakim, Alaa E and Alfattni, Ghada and Althobaiti, Hassan and Ikram, Mohammed and Alharbi, Amirah and Alsharif, Hussam and Alshamrani, Mazin and Alharbi, Ebtisam and others},
  journal={IEEE Access},
  year={2025},
  publisher={IEEE}
}

@inproceedings{jayasudha2025real,
  title={Real-Time Crowd Management and Crime Prevention System Using AI},
  author={Jayasudha, M and Chalappuram, Joel J and Edla, Bhargava Ram},
  booktitle={International Conference on Artificial Intelligence, Communication Technologies \& Smart Cities},
  pages={557--570},
  year={2025},
  organization={Springer}
}

@inproceedings{deng2019arcface,
  title={Arcface: Additive angular margin loss for deep face recognition},
  author={Deng, Jiankang and Guo, Jia and Xue, Niannan and Zafeiriou, Stefanos},
  booktitle={Proceedings of the IEEE/CVF conference on computer vision and pattern recognition},
  pages={4690--4699},
  year={2019}
}
 
\end{document}